\documentclass[11pt,a4paper]{article}

\usepackage{amsmath}
\usepackage{braket}
\usepackage{epigraph}
\usepackage{jheppub}
\usepackage{tensor}
\usepackage{enumerate}
\usepackage{slashed}
\usepackage{graphicx}
\usepackage{subfig}

\usepackage{feynmp}
\makeatletter
\def\endfmffile{%
  \fmfcmd{\p@rcent\space the end.^^J%
          end.^^J%
          endinput;}%
  \if@fmfio
    \immediate\closeout\@outfmf
  \fi
  \ifnum\pdfshellescape=\@ne
    \immediate\write18{mpost \thefmffile}%
  \fi}
\makeatother

\date{}

\title{Gravitational leptogenesis, C, CP and strong equivalence}

\author{Jamie I. McDonald and Graham M. Shore}
\affiliation{Department of Physics,\\
Swansea University,\\
Swansea, SA2 8PP,  UK.}

\emailAdd{g.m.shore@swansea.ac.uk}
\emailAdd{pymcdonald@swansea.ac.uk}

\abstract{The origin of matter-antimatter asymmetry is one of the most important
outstanding problems at the interface of particle physics and
cosmology. Gravitational leptogenesis (baryogenesis) provides a
possible mechanism through explicit couplings of spacetime curvature
to appropriate lepton (or baryon) currents. In this paper, the idea
that these strong equivalence principle violating interactions could
be generated automatically through quantum loop effects in curved
spacetime is explored, focusing on the realisation of the
discrete symmetries C, CP and CPT which must be broken to induce
matter-antimatter asymmetry. The related issue of quantum corrections
to the dispersion relation for neutrino propagation in curved
spacetime is considered within a fully covariant framework.
\vskip4cm}

\newpage

\notoc
\begin{document}
\maketitle

\section{Introduction}

The origin of matter-antimatter asymmetry in the universe remains one
of the outstanding problems at the interface of particle physics and
cosmology. In recent years, fresh impetus has been given to this issue
by the development of models where gravity plays a key role in the
symmetry breaking dynamics.

The motivation for looking at gravitational leptogenesis\footnote{In
  this paper we focus on leptogenesis, although all the theoretical
  development would transfer immediately to direct models of
  gravitational baryogenesis.  A lepton-antilepton asymmetry may also
  be transferred to a baryon-antibaryon asymmetry by the standard
  sphaleron mechanism at non-zero temperature \cite{Kuzmin:1985mm}.}
arises from a critical analysis of the Sakharov conditions
\cite{Sakharov} for the
generation of matter-antimatter asymmetry in cosmology. These require
models of lepto/baryogenesis to contain~(i) a mechanism for
lepton/baryon number violation, ~(ii) C and CP violation, ~(iii)
non-equilibrium dynamics. Subsequently, it was realised that in a
gravitational field, the final criterion may be replaced by an
effective violation of CPT symmetry \cite{Cohen:1987vi}. 
Models with C and CP-violating gravitational couplings introduced by
hand in a more or less well-motivated way have also been proposed to
address the second Sakharov condition (see {\it
  e.g.}~ref.\cite{Lambiase:2013haa} for a review). An important example is the
interaction $\partial_\mu R J^\mu$, where $J^\mu$ is the lepton
current, introduced in the model of ref.\cite{Davoudiasl:2004gf}. Here, the
time derivative of the Ricci scalar, $\dot{R}$, may act as a chemical
potential for lepton number, inducing a lepton-antilepton asymmetry at
non-zero temperature.

On the other hand, it is known that in quantum field theory in curved
spacetime, quantum loop effects induce effective violations of the
strong equivalence principle in the sense that the corresponding
effective Lagrangian contains interaction terms which depend
explicitly on the curvature, such as
$R_{\mu\nu\rho\sigma}F^{\mu\nu}F^{\rho\sigma}$
or $R_{\mu\nu}\bar\psi \gamma^\mu D^\nu\psi$
~\cite{Drummond:1979pp,Ohkuwa:1980jx}.
The question naturally arises whether we can use this mechanism of
radiatively-induced strong equivalence violation to automatically
generate curvature interactions relevant for leptogenesis. Moreover, a
better understanding of the mechanism -- especially of the role of the
discrete symmetries C, P and T and the key combinations CP and CPT --
should provide a guide to the properties of BSM models necessary for
gravitational leptogenesis to work. The purpose of this paper is to
develop a better theoretical understanding of these fundamental
issues.

We should be clear to distinguish two interpretations of
``symmetry breaking'' in this context. First there is the question of
whether the full terms arising in the effective Lagrangian,
including the curvature, are invariant under the symmetries of the
original theory; in particular, whether discrete symmetries present at
tree level may be violated by quantum loop effects. For example, we
may ask whether an interaction of the form $\partial_\mu R \bar\psi
\gamma^\mu \psi$, which is CP odd, can arise from a CP-conserving 
tree-level action. We refer to this as ``{\it anomalous symmetry
  breaking}''. For this, we require a careful discussion of the
realisation of discrete symmetries in a spinor theory in curved
spacetime.

However, at a given point in spacetime where the background curvature
takes a fixed value, the effective Lagrangian resembles the Lorentz
and CPT-violating actions proposed by Kostelecky and collaborators
\cite{Colladay:1998fq},
with the background field curvature playing the role of coupling constants for
potentially C, CP or CPT-violating operators. For example, the
Lorentz-violating Dirac action contains a term 
$a_\mu \bar\psi \gamma^\mu \psi$ where the coupling constant $a_\mu$
multiplies an operator which is CPT odd. It follows that in a fixed
background, the radiatively-induced curvature interactions may
{\it effectively} violate these discrete symmetries, giving rise to
phenomenological effects comparable to those in explicit Lorentz-violating
theories. We will refer to this as ``{\it environmental symmetry
  breaking}'' to emphasise the distinction.

Although we are primarily motivated by possible applications to
leptogenesis, the paper is also concerned with more general
theoretical issues related to quantum loop effects in spinor field
theories in curved spacetime, especially gravitational effects on 
neutrino propagation. In fact, these are closely related since any
gravity-induced change in the neutrino dispersion relation which
distinguishes between left and right-handed fermions would induce a
matter-antimatter asymmetry.

The paper is organised as follows. In section 2, we present a
fully covariant derivation of the dispersion relation for Dirac
fermions propagating in curved spacetime, in the framework of the
eikonal approximation. Establishing and understanding this requires a
careful treatment of the role of local inertial frames and the spin
connection, so we begin with an extended discussion of the geometry of
spinors in curved spacetime. This discussion sharpens our critique of
a number of proposals in the literature, in particular the suggestion
in refs.\cite{Singh:2003sp,Mukhopadhyay:2005gb,Debnath:2005wk} 
that leptogenesis can arise simply
through the coupling of neutrinos to the spin connection in the
tree-level Dirac action. 

Section 3 contains our main analysis of the effective violation of the
strong equivalence principle by quantum loops in the case of neutrinos
in the standard model. We determine the one-loop effective Lagrangian,
emphasising the importance of using a complete basis of hermitian
operators, by matching coefficients with explicit Feynman diagram 
calculations in a weak background field. This generalises earlier work
on neutrino propagation in curved spacetime by Ohkuwa \cite{Ohkuwa:1980jx}. 

We then analyse in some detail the discrete symmetry properties of the 
operators arising in the effective Lagrangian. We verify that in this model,
only CP even operators arise, respecting the symmetries of the
original tree-level action. There is no evidence, even in curved
spacetime, for an {\it anomalous} violation of C, CP or CPT symmetry
at the quantum loop level. In particular, a CP-violating interaction 
$\partial_{\mu} R ~\overline{\nu}_R \gamma^\mu \nu_L$ of the type required by the
Davoudiasl {\it et al.} model of leptogenesis \cite{Davoudiasl:2004gf} 
does not arise through
radiative corrections in the CP-conserving sector of the standard
model. We conclude that such curvature interactions would only 
arise in theories in which some CP violation is already present in
the original Lagrangian.

Applications of strong equivalence violating curvature interactions in
gravitational leptogenesis generally rely either on identifying an
interaction as analogous to a chemical potential for lepton number or
inferring a splitting in energy levels for particles and antiparticles through
dispersion relations. With this motivation, in section 4 we study in
some detail the dispersion relations arising from the operators which
appear in the effective Lagrangian of section 3, also including the
CP-violating operator described above. The occurrence of a hierarchy
of scales when curvature interactions are present requires some
generalisation of the eikonal approximation discussed in section
2. This also allows us to determine the quantum loop corrections to
the neutrino dispersion relation in the standard model, verifying the
result of \cite{Ohkuwa:1980jx} that the low-frequency phase velocity for
massless neutrinos is superluminal for backgrounds satisfying the
null-energy condition.

Finally, in section 5, we summarise our conclusions and discuss the
implication of the theoretical issues raised in our work to models
which attempt to generate matter-antimatter asymmetry through
gravitational leptogenesis.

\section{Inertial Frames and Spinors}

\subsection{Spinor Formalism in Curved Spacetime}

In this section we give a brief review of the main elements of the spinor
formalism in curved spacetime we need in the paper
(see \cite{Bir,Free} for a more complete account). In a general background, Lorentz transformations can only be realised locally in the tangent plane at each spacetime point. This is achieved by introducing an orthonormal basis $\lbrace e_\mu^{(a)} \rbrace$ satisfying
\begin{equation}\label{mink}
\eta_{ab} = e_{(a)}^\mu e_{(b)}^\nu g_{\mu \nu}.
\end{equation}
Here, Greek indices label coordinate basis components and Latin indices label the different vierbein basis vectors $a=0,1,2,3$. Lorentz transformations are defined as any transformation of the vierbein
\begin{equation}
e^{\mu}_{(a)}(x) \rightarrow e'^{\mu}_{(a)}(x)=\tensor{\Lambda(x)}{_a^b}e^{\mu}_{(b)}(x), 
\end{equation}
with
\begin{equation}
\Lambda(x) = \exp \left(-\frac{1}{2}\Omega_{ab}(x) M^{ab}\right)
\end{equation}
which preserves the relation (\ref{mink}), where $M^{ab}$ form a basis
for the \textit{fundamental} representation of the Lorentz algebra. 
Thus the vierbein provides a rectangular frame on which one can perform local boosts and rotations. 

Now that we have formulated Lorentz transformations, we can introduce
particles in the spin-1/2 representation of the Lorentz group. 
Since the Lorentz transformations are local, this necessitates the
introduction 
of a gauge connection or \textit{spin-connection} $\omega_\mu$ in the spinor representation of the Lorentz algebra 
\begin{equation}
\omega_\mu = \omega_\mu^{ab}\sigma_{ab},
\end{equation}
where $\sigma^{ab} = \frac{i}{2}[\gamma^a,\gamma^b]$. The covariant derivative is then defined by
\begin{equation}
D_\mu \psi = \left( \partial_\mu - \frac{i}{4} \omega_\mu^{ab} \sigma_{ab} \right) \psi,
\end{equation}
which transforms in the same way as $\psi$ under SO(1,3) transformations:
\begin{equation}
\psi(x) \rightarrow \exp\left(-\frac{i}{2} \Omega_{ab}(x) \sigma^{ab}\right) \psi(x) \equiv D[\Lambda(x)]\psi(x),
\end{equation}
provided that the spin connection transforms as
\begin{equation}
\omega^\mu_{ab} \rightarrow\tensor{\Lambda}{_a ^ c} \tensor{\Lambda}{_b ^d} \omega^\mu_{cd} +  \tensor{\Lambda}{_a ^c} \partial_\mu \Lambda_{bc} 
\end{equation}
Defining $\gamma^\mu = e^{\mu}_{(a)}\gamma^a$, one can construct a Lorentz invariant Dirac action 
\begin{equation}
\mathcal{L} = \bar{\psi} \left( i \gamma^\mu D_\mu - m \right) \psi. 
\end{equation}
One can find a relation between the spin-connection $\omega^{ab}_\mu$, vierbein and Christoffel symbols in the following way. Consider a vector $X$. We have that
\begin{align}
\nabla X & = \left(\nabla_\mu X^\nu \right) dx^\mu \otimes \partial_\nu \\ \label{coord}
& = \left( \partial_\mu X^\nu + \Gamma^\nu_{\lambda \mu}X^\lambda \right)dx^\mu \otimes \partial_\nu.
\end{align}
We can also write the derivative in the vierbein basis $\nabla X = \left( \nabla_\mu X^a \right) dx^\mu \otimes e_{(a)}$, which after a little algebra leads to the relation
\begin{equation}\label{nonhol}
\nabla X = \left( \partial_\mu X^\nu + e^{\nu}_{(a)} \partial_\mu e^{(a)}_\lambda X^\lambda + e^\nu_{(a)}e^{(b)}_\lambda  \tensor{(\omega_\mu)}{^a_b} X^\lambda \right) dx^\mu \otimes \partial_\nu.
\end{equation}
In order for the expressions (\ref{coord}) and (\ref{nonhol}) to agree, we must have 
\begin{equation}
\tensor{\left( \omega_\mu \right)}{^a _ b}  = e^{(a)}_\nu \left( \partial_\mu e_{(b)}^\nu  + \Gamma^\nu_{\sigma \mu} e^\sigma_{(b)}\right). 
\label{spincon}
\end{equation}
With regard to notation, we will use $D_\mu$ to denote the derivative which is both a $GL_4(\mathbb{R})$ and $SO(1,3)$ tensor. Its action on the vierbein is defined by
\begin{equation}
D_\mu e_\nu^{(a)}= \partial_\mu e^{(a)}_\nu - \Gamma_{\nu \mu}^{\sigma} e_\sigma^{(a)} - \tensor{(\omega_\mu)}{^a _ b} e^{(b)}_\nu,
\end{equation}
so that (\ref{spincon}) is equivalent to the condition
\begin{equation}
D_\mu e_\nu^{(a)} =0.
\end{equation}
We will use $\nabla_\mu e_\nu^{a}$ to denote the derivative which is a $GL_4(\mathbb{R})$ tensor, but not a $SO(1,3)$ tensor:
\begin{equation}
\nabla_\mu e^\nu_{(a)}  = \partial_\mu e^\nu_{(a)} + \Gamma^\nu_{\sigma \mu}e^{\sigma}_{(a)},
\end{equation} 
This allows the relation (\ref{spincon}) to be written in a more compact form as 
\begin{equation}\label{spin}
\tensor{\left( \omega_\mu \right)}{^a _ b}  = e^{(a)}_\nu \nabla_\mu  e_{(b)}^\nu. 
\end{equation}
In the next section we introduce the concept of a non-accelerating
vierbein frame, and use this together with the relation (\ref{spin}) to show
how the spin connection must vanish at the origin of such a frame and
hence that the Dirac equation satisfies the strong equivalence principle.

\subsection{Inertial Frames}

In flat space, the Cartesian tetrad has constant components along any
curve, and thus defines an inertial tetrad throughout spacetime. In
curved space, an ``inertial" tetrad can only be defined in the
neighbourhood of a specific reference point corresponding to the
inertial observer. 
The covariant generalisation of ``constant components" along a curve is parallel transport.

\begin{figure}[h]
\centering
\includegraphics[trim=0cm 10cm 0cm 5cm, clip=true, scale=0.3]{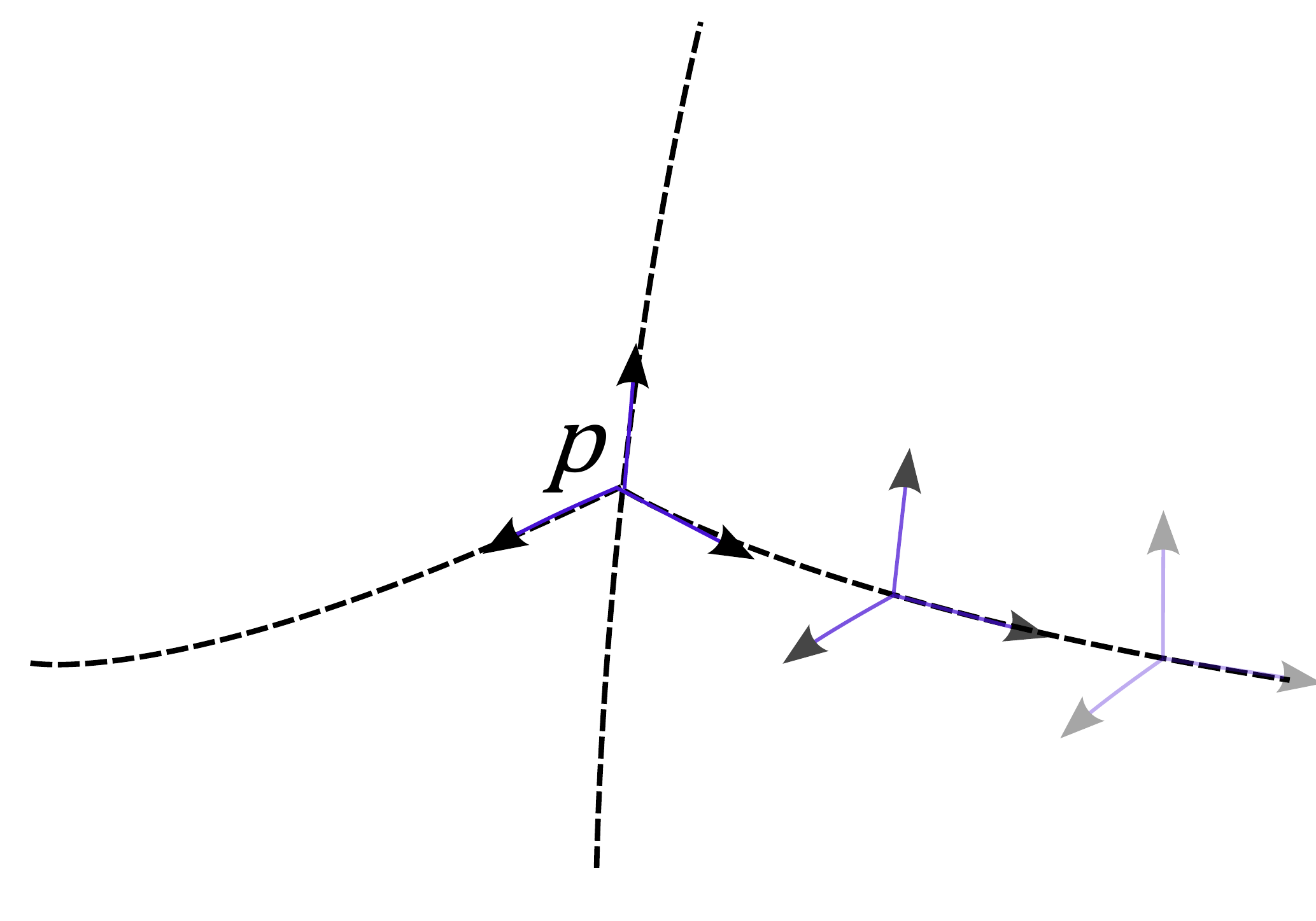}
\caption{The vierbein is defined in the neighbourhood of $p$ by parallel transport along all curves emanating from $p$.}
\label{transport}
\end{figure}

We construct an \textit{inertial frame} about a given point $p$ as
follows. Choose any orthonormal tetrad $\lbrace e_{(a)}^\mu
(p)\rbrace$ at $p$ and define the vierbein in the neighbourhood of $p$
by parallel transport of the vierbein along every curve emanating from $p$ (see figure \ref{transport}). It is now easy to see that the spin connection must vanish at $p$.  Pick any coordinate chart $x^\mu$ in the neighbourhood of the point $p$, and let $V^\mu$ be the tangent vector of any curve through $p$. Then we have
\begin{equation}
V^\mu \omega_\mu^{cd}= e^{(c)}_\nu  V^\mu \nabla_\mu e^{(d) \nu},
\end{equation}
but the parallel transport condition means that\footnote{Notice the derivative here gives an SO(1,3) gauge fixing, and should not be confused with the one in the equation $D_\mu e_{\nu}^{(a)}=0$, which is SO(1,3) invariant.} $V^\mu \nabla_\mu e^{(d)}=0$ at $p$ and since $V_\mu$ is arbitrary it follows that $\omega_\mu$ vanishes at $p$. The existence of a local inertial frame is guaranteed by the assumption that the spacetime is Riemannian, the mathematical realisation of the weak equivalence principle.

The physical interpretation is that parallel transport ensures the vierbein, which may be thought of as a set of measuring rods, is not accelerating as it approaches $p$. For a given direction specified by a curve through $p$ with tangent vector $V^\mu$ , one can define an acceleration 4-vector $a^\mu_{(a)}$ for each vierbein component
\begin{equation}
V \cdot \nabla e^{(a)}_\mu = a_\mu^{(a)}.
\end{equation}
It is then easy to see that our prescription simply imposes the condition that the 4-acceleration of any measuring rod is zero in all directions approaching $p$. Put another way, it means that only observers whose ``measuring rods" are accelerating will measure the spin connection at $p$. 

Another way to understand the parallel transport condition is to set
up Riemann normal coordinates centered on $p$. In these coordinates
$\Gamma_\mu^{\sigma \rho} (p)=0$ 
and so the condition $\nabla  e^{(a)} (p)= 0$ is simply
\begin{equation}
\partial_\mu e^{(a)}(p) =0.
\end{equation}
In other words, observers with inertial coordinates perceive the inertial vierbein to have constant components in an infinitesimal neighbourhood of $p$. It is now easy to see that for an inertial tetrad the Dirac equation satisfies
\begin{equation}
\mathcal{L}=\bar{\psi}\left( i \slashed{D} -m \right) \psi \rightarrow \bar{\psi}\left( i \slashed{\partial} -m \right) \psi, 
\end{equation}
at $p$. This satisfies the strong equivalence principle, {\it viz.}~it
reduces to its special relativistic form in an inertial frame. As we
have seen, this corresponds to the requirement that the curved space Lagrangian
involves only the connection, with no explicit curvature terms. We will see how this is affected by quantum corrections later in the paper.

We should also mention that one can define an inertial vierbein along a curve $\gamma$ associated to a geodesic observer with tangent vector $u^\mu$ by choosing $e^{(0)}=u$ (with the normalisation $u^2=1$) and defining $e^{(i)}$ in the neighborhood by parallel transport along spacelike geodesics normal to $\gamma$. This fixes the SO(1,3) gauge in the normal convex neighbourhood of $\gamma$. One can then define an inertial set of coordinates $x^{\hat{\mu}}$ around $\gamma$ in which the Christoffel symbols vanish along $\gamma$, i.e. $\Gamma_{\hat{\mu} \hat{\nu}}^{\hat{\rho}}(\gamma)=0$. Physically this corresponds to a freely falling observer carrying a gyroscopic set of measuring rods. The full mathematical formulation of this concept is \textit{Fermi normal coordinates} as discussed at length in \cite{Poisson:2011nh}.

\subsection{Particle Propagation and the Dirac Equation}

In curved spacetime, familiar Minkowski space concepts such as
particle momenta and trajectories, spin states and dispersion
relations are no longer directly applicable and their generalisation
requires a subtle and careful analysis of the Dirac equation and its
relation to the underlying geometry. 
Our starting point is the Dirac equation in curved spacetime 
\begin{equation}\label{Dirac}
\left( i \slashed{D} - m\right)\psi =0.
\end{equation}
In flat space, particles are identified with plane waves, but the
curved space Dirac equation will not in general admit such
solutions. However, in a kinematical regime where the curvature scale
is relatively long compared to the wavelength, we can find solutions
which locally resemble plane waves and thus exhibit particle-like 
properties. 

\begin{figure}[ht!]
\centering
\includegraphics[trim=3cm 12cm 0cm 7cm, clip=true, scale=0.38]{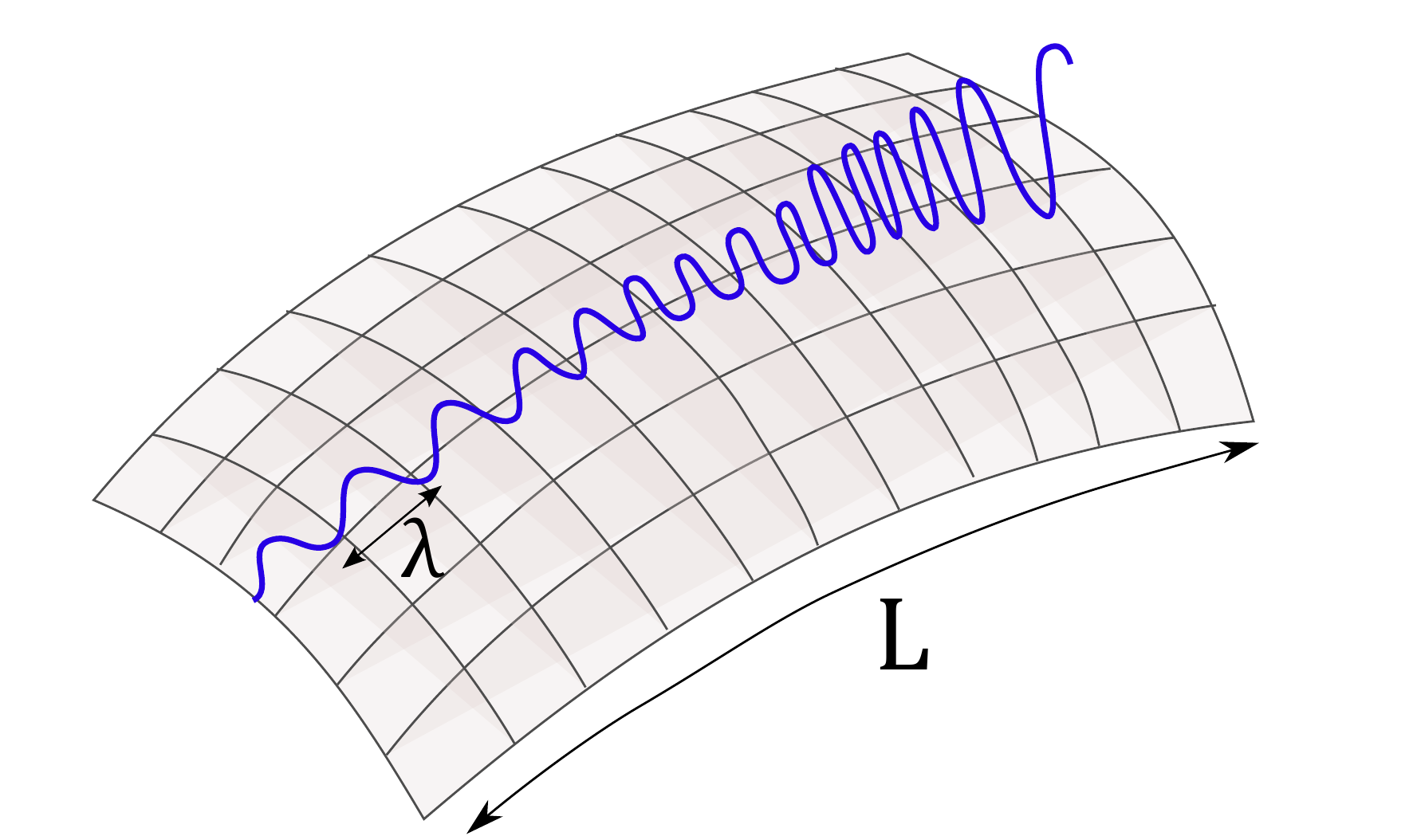}
\caption{An illustration of the eikonal approximation. The wavelength
  $\lambda$ 
is much less than typical curvature scales $L$, so that solutions 
locally resemble plane waves.}
\label{pic}
\end{figure}

To construct these quasi-plane wave solutions, we use an
eikonal approach familiar from geometric optics in curved space
\cite{MTW, Schneider}. These solutions are characterised by a
wavelength $\lambda = 1/|\underline{k}|  \ll L$ where 
$k_\mu$ is the wave-vector and $L$ is the scale over which
the amplitude varies (see figure \ref{pic}).

Two other scales enter the analysis -- the Compton wavelength of the
Dirac particle $\lambda_c = 1/m$ and the curvature scale 
$1/\sqrt{\cal{R}}$, where  ${\cal{R}}$ represents the size of a
typical curvature tensor component. We need to decide from the outset
how these relate to the hierarchy of scales $\lambda, L$. 
Since in the flat-space limit we want to recover the standard
dispersion relation $k^2 - m^2 = 0$, we should clearly take $\lambda_c
\ll L$. We also identify $1/\sqrt{\cal{R}}$ with $L$,
since it is the background gravitational field that determines the
scale over which the amplitude of the quasi-plane waves will vary.
If we were to take the curvature scale $1/\sqrt{\cal{R}}$ comparable
to $\lambda_c$ or $\lambda$, the solutions would no longer resemble
plane waves and we would lose any interpretation in terms of
particle states carrying definite momenta.

In the eikonal approximation, we split the solutions into a
rapidly-varying phase $\Theta(x)$ and a slowly-varying amplitude
${\cal A}^{(s)}$ ($s$ = 1,2) multiplying basis spinors $u^{(s)}$ (or $v^{(s)}$ for
the corresponding antiparticle solutions), {\it i.e.}
\begin{equation}\label{eika}
\psi(x)= {\cal A}^{(s)}(x) u^{(s)}(x)  e^{-i \Theta(x)/\epsilon},
\end{equation}
where 
\begin{equation}\label{eikb}
{\cal A}^{(s)}(x) = a^{(s)}(x)  - i b^{(s)}(x) \epsilon - c^{(s)}(x)
\epsilon^2 + \ldots 
\end{equation}
The book-keeping parameter $\epsilon$ (\cite{MTW, Schneider}), which is
finally set to 1, identifies the order of the associated quantities in
powers of the parameter $\lambda/L$. We then solve the equation order
by order in an expansion in powers of $\epsilon$.  To implement the
condition $\lambda_c \ll L$ automatically, we should also take the mass
$m\rightarrow m/\epsilon$ in the Dirac equation (\ref{Dirac}).

The simplest way to derive the key results, and to compare with
previous analyses of the Maxwell equation and photon
propagation \cite{Shore:2003jx}, is to square the Dirac equation and consider
the wave equation 
\begin{equation}\label{Dir2}
\left( D^2 + \frac{m^2}{\epsilon^2} - \frac{1}{4}R \right) \psi =0
\end{equation}
Notice that the Ricci scalar arises from the identity
\begin{align}
\left[D_\mu, D_\nu \right]\psi  = \frac{1}{4}R_{\mu \nu \rho \sigma} \gamma^\rho \gamma^\sigma \psi
\end{align}
and gamma matrix manipulations. Now insert the eikonal ansatz (\ref{eika}),(\ref{eikb}) into (\ref{Dir2}) and identify $k_\mu
= \partial_\mu \Theta$ as the wave-vector, which is orthogonal 
to the wavefronts of constant phase $\Theta$.  Collecting terms of the
same order in $\epsilon$, we find a solution by satisfying the
following equations sequentially for $k^2$, $a^{(s)}$ and $b^{(s)}$:
\begin{align}
O(1/\epsilon^2) : &&   \left(k^2 - m^2 \right) a\cdot u =0 \label{k2}\\
O(1/\epsilon): && -\left[ 2k \cdot D + \left(D \cdot k\right) \right] a\cdot u+\left(k^2 - m^2 \right) b\cdot u =0 \label{aa1}\\
O(1): &&  \left(D^2 - \frac{1}{4}R \right) a \cdot u - \left[ 2 k
  \cdot D + \left( D \cdot k \right) \right]b\cdot u 
+ \left(k^2 - m^2 \right) c\cdot u =0 \label{curv},
\end{align}
where $a\cdot u$ is a shorthand for $a^{(s)} u^{(s)}$ with no sum on $s$. Equation (\ref{k2}) recovers the expected dispersion relation 
\begin{equation}\label{disp0}
k^2-m^2=0.
\end{equation}
To understand the geometric significance of the second equation, 
consider the congruence defined by the tangent 
vectors $\hat k^\mu = g^{\mu\nu}k_\nu/m$. These curves, which 
are timelike geodesics, are identified as the particle
trajectories in curved space. The geodesic property follows 
immediately from (\ref{disp0}) by  taking a covariant derivative, and using the identity 
$\nabla_{[\mu} k_{\nu]} =0$, giving the geodesic equation 
\begin{equation}
k^\mu \nabla_\mu k^\nu =0.
\end{equation}
Defining $\Omega_{\mu\nu} = \nabla_\mu \hat{k}_\nu$, we may identify
the optical scalars $\theta$ (expansion), $\sigma$ (shear) and
$\omega$ (twist) of the congruence as
\begin{equation}
\Omega_{\mu \nu} = \frac{1}{3}\theta P_{\mu \nu} + {\sigma}_{\mu \nu} + {\omega}_{\mu \nu},
\end{equation}
where we define the projection operator for the hypersurfaces of constant phase by
\begin{equation}
P_{\mu \nu} = g_{\mu \nu} - \hat{k}_\mu \hat{k}_\nu,
\end{equation}
and where
\begin{align}
\theta & = \tensor{\Omega}{^\mu_\mu}, \\
{\sigma}_{\mu \nu} & = \Omega_{(\mu \nu)} - \frac{1}{3}\theta P_{\mu \nu},\\
\omega_{\mu \nu} & = \Omega_{[\mu \nu]} .
\end{align}
The congruence is twist-free (and therefore surface-forming) by virtue of the fact that $k_\mu$ is a gradient, so $\nabla_{[\mu} k_{\nu]} =0$. The divergence measures the rate of expansion of the congruence,  $\omega$ measures the tendency of the congruence to twist and the shear $\sigma$ corresponds to geodesics moving apart in one direction and together in the orthogonal direction, whilst preserving the cross-sectional area.

Given $k^2=m^2$, it follows that the $O(1/\epsilon)$ equation
(\ref{aa1}) becomes
\begin{equation}\label{ampev}
\left[ 2k \cdot D + \left(D \cdot k\right) \right] a^{(s)} u^{(s)} =0.
\end{equation}
If we choose a solution in which the basis spinors $u^{(s)}$ are paralelly transported:
\begin{equation}\label{para}
k \cdot D u^{(s)}=0,
\end{equation}
the equation (\ref{ampev}) becomes
\begin{equation}\label{expan}
\hat{k} \cdot D a^{(s)} = - \frac{1}{2}\theta a^{(s)},
\end{equation}
which shows that at leading order the amplitude is governed
geometrically by the expansion rate of the geodesic congruence. 
The $O(1)$ equation (\ref{curv}) then determines the sub-leading
amplitude correction $b^{(s)}$ in terms of $a^{(s)}$, and so on.

So far, our results are independent of a particular choice of frame.
Now, choose a particular timelike geodesic $\gamma$ within the
congruence and imagine
a co-moving observer freely-falling with the particle along this
trajectory, measuring the evolution of its spin polarization. 
The observer is equipped with an inertial vierbein, where we identify
the timelike vierbein component $e^\mu_{(0)}$ with $\hat k^\mu$ and
demand that the spatial vierbein vectors $e^\mu_{(i)}$ are parallel
transported in the neighbourhood of  $\gamma$. In this inertial frame
the connection vanishes along the trajectory,
\begin{equation}
\omega^\mu_{ab} \vert_\gamma =0
\end{equation}
Equations (\ref{para}), (\ref{expan}) then simplify and we have 
\begin{equation}
k^\mu \partial_\mu u^{(s)}\vert_\gamma = 0
\end{equation}
showing that in this inertial frame the basis spinors are constant
along the trajectory, while the amplitude satisfies
\begin{equation}
k^\mu \partial_\mu a^{(s)} \vert_\gamma = - \frac{1}{2}\theta a^{(s)}
\end{equation}

Several key points need to be emphasised here:

\begin{enumerate}[(i)]
\item in the eikonal approximation, we recover a particle
interpretation even in curved space, with Dirac particles propagating
along timelike geodesics;
\item the connection does {\it not} appear in the dispersion
relation which, within the eikonal approximation, is identical to its
flat space form consistent with the weak and strong equivalence principles;

\item  the spin polarisation is parallel propagated along the
trajectory, and is constant viewed in an inertial frame;

\item  the wave amplitude (particle density) is governed by
the expansion optical scalar $\theta$ of the associated geodesic
congruence;

\item  the curvature only affects the amplitude, not the
dispersion relation, and only at higher order in the eikonal
approximation as given by (\ref{curv}).
\end{enumerate}

While this reveals the essential physics, and allows an easy
comparison with photon propagation, a more rigorous treatment demands
that we solve the Dirac equation itself in this framework
rather than just the associated wave equation (\ref{Dir2}).
This was first carried out by Audretsch \cite{Audretsch:1981wf} and we
present a simplified version of his analysis here.

Acting with the Dirac operator on the eikonal ansatz (\ref{eik1}) and collecting
terms of the same order in $\epsilon$ as before, this time we find
\begin{align}
O(1/\epsilon): & \qquad  \left( \slashed{k}-m \right) a \cdot u =0 \label{eik1}\\
O(1): & \qquad \slashed{D}a \cdot u - \left(\slashed{k} - m \right)b\cdot u =0 \label{eik2}\\
O(\epsilon): & \qquad \slashed{D} b \cdot u -\left( \slashed{k} - m \right) c \cdot u =0
\end{align}
The first equation is now an algebraic matrix equation and the
existence of a non-trivial solution requires 
\begin{equation}
\det \left(\slashed{k}-m \right) =0
\end{equation}
from which we recover the original dispersion relation $k^2 - m^2=0$.
Taking the hermitian conjugate of (\ref{eik1}), we see from left multiplying (\ref{eik2}) by $ \bar{u}^{(s)}$ that
\begin{equation}\label{amp}
\bar{u}^{(s)} \slashed{D} \left( a^{(s)} u^{(s)} \right) =0
\end{equation}
As before we want to choose basis spinors $u^{(s)}$ which satisfy the normalisation
\begin{equation}
\bar{u}^{(r)} \gamma^\mu u^{(s)} = \hat{k}^\mu \delta^{rs}
\label{norm}
\end{equation}
and the parallel propagation property
\begin{equation}
\hat{k}^\mu D_\mu u^{(s)} =0.
\end{equation} 
One possible choice is \footnote{It can be most easily checked that
  these satisfy these 
two conditions by evaluating in the rest frame of the particle where
$k^{(i)}=0$ 
and using the fact they are SO(1,3) tensor identities.}
\begin{equation}
u^{(1)}  = \left( \frac{E+m}{2m} \right)^{1/2}
\left(
\begin{array}{c}
1\\
\\
0\\
\\
\frac{k^{(3)}}{E+m}\\
\\
\frac{k^{(1)}+ik^{(2)}}{E+m}
\end{array}
\right),
\qquad
u^{(2)}   = \left( \frac{E+m}{2m} \right)^{1/2} \left(
\begin{array}{c}
0\\
\\
1\\
\\
\frac{k^{(1)}-ik^{(2)}}{E+m}\\
\\
-\frac{k^{(3)}}{E+m}
\end{array}
\right),
\end{equation}
with
\begin{equation}
E = k^\mu e^{(0)}_\mu, \qquad k^{(i)} = k^\mu e^{(i)}_\mu.
\end{equation}
Taking a covariant derivative of (\ref{norm}) gives
\begin{equation}
\bar{u}^{(r)} \slashed{D} u^{(s)} = \frac{1}{2}\theta \delta^{rs},
\label{expans}
\end{equation}
so that expanding (\ref{amp}) as
\begin{equation}
\bar{u}^{(r)} \gamma^\mu u^{(s)} D_\mu a^{(s)} + a^{(s)}\bar{u}^{(r)}\slashed{D}u^{(s)} =0
\end{equation}
and substituting (\ref{norm}) and  (\ref{expans}) gives the evolution of the amplitude as before:
\begin{equation}
k^\mu D_\mu a^{(s) } = -\frac{1}{2}\theta a^{(s)}.
\end{equation}

Finally, we should perhaps emphasise that while this shows that the 
dispersion relation is independent of the spin connection in the free Dirac
Lagrangian, it does {\it not} mean that the gravitational field has 
no influence the dynamics of the fermion's spin. Indeed, Audretsch
\cite{Audretsch:1981wf} has analysed the propagation of the fermion current
$\bar\psi \gamma^\mu \psi$ in more detail, using a Gordon
decomposition into a `convection current' $ -i \bar\psi
\overleftrightarrow{D^\mu} \psi$ and a `spin magnetisation current'
$(\bar\psi \sigma^{\mu\nu}\psi)_{;\nu}$.  At leading order in the eikonal expansion,
the convection current follows the timelike geodesic defined
by $\hat k^\mu$ (at higher order there are tidal curvature
corrections), while the spin motion is defined by parallel propagation.

\subsection{Dispersion Relations and Covariance}

This analysis allows us to understand better a number of proposals in
the literature aimed at exploiting a background gravitational field to
induce baryo/leptogenesis through modified dispersion relations.

In a series of papers, Mukhopadhay and others 
\cite{Singh:2003sp,Mukhopadhyay:2005gb,Debnath:2005wk} 
have looked for consequences of rewriting the Dirac Lagrangian in the
suggestive form
\begin{align}
\mathcal{L} &=\bar{\psi}\left( i \slashed{D} -m \right) \psi \\
&=  \bar{\psi}\left( i \gamma^a\partial_a -m \right) \psi 
+ B_a \bar{\psi} \gamma^a \gamma^5 \psi \label{chem}
\end{align}
in a vierbein frame, where
\begin{equation}
B^d = \epsilon^{abcd} \left( \omega_{bc} \right)_a,
\end{equation}
and where $\left( \omega_{bc} \right)_a = e_{(a)}^\mu
\left(\omega_{bc}\right)_\mu$ 
is the projection of the spin connection onto the vierbein basis. 
Since $\bar{\psi}\gamma^a \gamma^5 \psi$ is the spin current, it
appears that in a frame where $B_a$ is constant, this term acts as a
chemical potential which in a theory of neutrinos would induce a 
particle-antiparticle asymmetry.
To establish this, it is claimed
\cite{Singh:2003sp,Mukhopadhyay:2005gb,
Debnath:2005wk,Lambiase:2006md,Lambiase:2011by,Lambiase:2013haa} 
that the dispersion relation can be
obtained from (\ref{chem}) by considering plane wave solutions of the 
form\footnote{In fact, this solution has no real meaning in general
relativity as $p\cdot x =p^\mu x_\mu$, or $p^a x_a$, is not a well defined object 
except in Minkowski space. 
In relativity, $p^\mu$ lives in the tangent plane, but $x^\mu$ is just 
an element of the coordinate chart (not a vector) and so it makes no sense to define 
an ``inner product" between the two.}
$\exp(ip\cdot x)$ giving rise to
\begin{equation}\label{split}
(p^a \pm B^a)^2  = m^2
\end{equation}
for the left and right-handed particles respectively.

However, as we have seen, the true dispersion relations are
established in covariant form $p^2 = m^2$. In contrast,
equation (\ref{split}) is not covariant, since $B_a$ does not transform as
a tensor under the SO(1,3) Lorentz transformations
$e_{(a)} \rightarrow \tensor{\Lambda}{_a ^b}e_{(b)}$, but rather as
\begin{equation}
B^a \rightarrow \tensor{\Lambda}{^a _b}B^b +  \tensor{\Lambda}{^a _b} 
\tensor{\Lambda}{^c _d} \varepsilon^{bgdf}\partial_f \Lambda_{c g}
\end{equation}
Another way to see $B^a$ is not an SO(1,3) tensor is to note that we 
can always make it vanish at a point by $p$ by choosing an inertial 
vierbein there. Since tensors are either always zero at a point or 
never zero there, it cannot be a Lorentz tensor.

We conclude, therefore, that equation (\ref{split}) is {\it not} a valid
dispersion relation and has no consequence for gravitational
leptogenesis. More generally, any leptogenesis model which requires 
the non-vanishing of the spin connection (e.g. treating $B_a$ as a chemical potential), 
depends on working in a non-inertial, accelerating vierbein where $B_a \neq 0$. But this is giving
information purely on the nature of the acceleration, not revealing
the intrinsic covariant physics. While in some situations it is
appropriate to consider non-inertial observers, for cosmological
applications the appropriate 
frame in which to measure lepton density is that of an inertial
comoving observer. 
Thus the free Dirac Lagrangian in curved
spacetime does not give rise to gravitational leptogenesis. 

We should point out, however, that our conclusions apply to Riemannian
spacetimes, where the weak equivalence principle holds and the
connection vanishes in local inertial frames. Non-Riemannian geometry,
spacetimes with torsion, or string backgrounds with additional
antisymmetric and dilaton background fields in addition to the metric
\cite{Ellis:2013gca} 
may still present interesting generalisations of the picture presented
in the last section.

\section{Radiatively Induced SEP violation}

We now turn to the main topic of this paper, radiatively induced
strong equivalence breaking and the realisation of discrete
symmetries.
Once again, our main focus is on potential applications to
gravitational leptogenesis. We are therefore especially interested in
the automatic generation by quantum loop effects of operators such as
$\partial_\mu R \bar\psi \gamma^\mu \psi$ introduced by hand in the
model of Davoudiasl {\it et al.}~\cite{Davoudiasl:2004gf} as a C and
CP-violating source of
matter-antimatter symmetry.

The essential physics behind this effective violation of the strong
equivalence principle is readily understood. At the quantum loop
level, a particle no longer propagates as a point-like
object but is screened by the virtual cloud of particles appearing in
its self-energy Feynman diagram 
\cite{Drummond:1979pp,Ohkuwa:1980jx,Hollowood:2010bd,Hollowood:2011yh}.
As a result, it acquires an effective size characterised by the
Compton wavelength of the virtual particles, causing it to experience 
gravitational tidal forces through its coupling to the background curvature. 
Particle propagation at the quantum loop level is therefore
described by a mean field $\psi$  whose dynamics are described by 
the effective action $\Gamma$, 
which gives rise to the quantum-corrected equations of motion 
\begin{equation}
\frac{\delta \Gamma}{\delta \bar{\psi}} =0.
\end{equation}

As discussed above, many models of gravitational leptogenesis consider 
interactions of neutrinos with background curvature 
\cite{Davoudiasl:2004gf,Lambiase:2006md,Lambiase:2011by,Lambiase:2013haa}. 
It is therefore interesting to perform a thorough investigation of the effective
action for neutrinos propagating in a gravitational background. 
In this section, we examine the effect of quantum loops on the
neutrino dispersion relation, and give a careful discussion of C, P
and CP symmetries of the 1-loop effective action. 

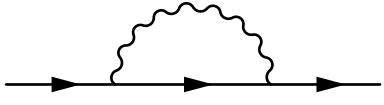
\begin{figure}[ht!]
\centering
\begin{fmffile}{sunset}
\begin{fmfgraph*}(140,60)
\fmfleft{i}
\fmfright{o}
\fmflabel{$\nu$}{i}
\fmflabel{$\nu$}{o}
\fmf{fermion,tension=2}{i,v1}
\fmf{fermion,tension=2}{v2,o}
\fmf{photon,label=$Z,,W$,left,tension=0.4}{v1,v2}
\fmf{fermion,label=$\nu,, e^-$}{v1,v2}
\end{fmfgraph*}
\end{fmffile}
\caption{Neutrino Self Energies}
\label{loop}
\end{figure}

Since we are interested in the propagation of neutrinos, 
we consider processes of the form shown in figure \ref{loop}.

\noindent The relevant parts of the SM lagrangian, with massless
neutrinos, are:

\begin{align}
\mathcal{L}_{EW}= \sqrt{-g} \Big[ &-\frac{1}{4} g^{\mu \rho}g^{\nu
  \sigma}
\left( 2W^-_{\mu \nu}W^+_{\rho \sigma} + Z_{\mu \nu}Z_{\rho \sigma}
\right) \nonumber\\ 
& + g^{\mu \nu} \left( M_W^2 W_\mu^-W^+_\nu + \frac{1}{2} M_Z^2 Z_\mu Z_\nu \right) \nonumber\\ 
& + \bar{e}\left( i {\gamma}\cdot D - m_e \right)e+ 
\bar{\nu}_R  i \gamma\cdot D  \nu_L \nonumber\\ 
& \frac{g}{\sqrt{2}} \left( \bar{\nu}_R \gamma \cdot W^+ e_L +
  \mbox{h.c} \right)
- \frac{g}{2 \cos
\theta_W}\bar{\nu}_R 
\gamma\cdot Z \nu_L \Big]  \label{EW},
\end{align}
where
\begin{equation}
\cos^2 \theta_W = \frac{g^2}{g'^2 + g^2}, \qquad m_W = \cos \theta_W m_Z 
\end{equation}
and $g$ and $g'$ are SU(2) and U(1) gauge couplings
respectively and $W_{\mu\nu}$, $Z_{\mu\nu}$ are the $W$ and $Z$ field
strengths. 
Since there are no infrared divergences in the relevant
Feynman diagrams and the electron mass contributes only at
$O(m_e/m_W)$, it has no qualitative effect on our analysis
and may be neglected.

\subsection{Hermiticity}\label{herm}

Since we are interested in the free propagation of neutrinos, we need
only consider those parts of the effective action quadratic in the
mean neutrino field, so that the equations of motion are linear in
$\nu$. We construct the effective action for operators up to third
order in derivatives by demanding that it is consistent with general
covariance and local Lorentz invariance. This is achieved by
constructing all possible neutrino bilinears from the 
contraction of curvature tensors $R, R_{\mu \nu}$ etc.~with gamma
matrices and neutrino spinors up to the required number of
derivatives. Up to third order in derivatives the following set of
operators covers all possible combinations:\footnote{With only
  left-handed neutrinos, there are no dimension 5 operators, since
  operators of the form $R \bar{\nu}_R\nu_L$ or  $\bar{\nu}_R D^2
  \nu_L$ vanish trivially. }
\begin{equation}
\partial_\mu R \bar{\nu}_R \gamma^\mu \nu_L, \quad R \bar{\nu}_R\slashed{D}\nu_L, \quad  R_{\mu \nu}\bar{\nu}_R \gamma^\mu D^\nu\nu_L, \quad \bar{\nu}_R D^2 \slashed{D}\nu_L, \quad \bar{\nu}_R\slashed{D} D^2 \nu_L, \quad  \bar{\nu}_R D_\mu \slashed{D}D^\mu \nu_L
\label{2nd},
\end{equation}
where
\begin{equation}
\nu_L = \frac{1-\gamma_5}{2}\nu, \qquad \bar{\nu}_R = \bar{\nu}\frac{1+\gamma_5}{2}.
\end{equation}
One might think that it is possible to construct operators from the
Riemann tensor $R_{\mu \nu \rho \sigma}$.
However the only possible combinations up to third order in derivatives are of the form
\begin{equation}
R_{\mu \nu \rho \sigma} \bar{\nu}_R\gamma^\mu \gamma^\nu \gamma^\rho D^\sigma \nu_L, 
\end{equation}
and so on, since they must involve only an odd number of gamma
matrices,
but using the Dirac algebra, and in particular the identities $R_{\mu [\nu \rho \sigma]}=0$ and
\begin{equation}
\gamma^\mu \gamma^\nu \gamma^\lambda = g^{\mu \nu}\gamma^\lambda + g^{\nu \lambda}\gamma^\mu - g^{\mu \lambda}\gamma^\nu - i \varepsilon^{\sigma \mu \nu \lambda}\gamma_\sigma \gamma_5, 
\end{equation}
it is straighforward to show that these reduce to linear combinations of the operators in  (\ref{2nd}). Finally using the identity
\begin{equation}
\left[D_\mu, D_\nu \right]\nu_L = \frac{1}{4}R_{\mu \nu \rho \sigma} \gamma^\rho \gamma^\sigma \nu_L
\end{equation}
it can be shown that the final two terms in (\ref{2nd}) give a contribution to the action which is a linear combination of other operators:
\begin{align}
\int d^4 x  \sqrt{-g} \left(\bar{\nu}_R D_\mu \slashed{D} D^\mu \nu_L \right) &= \int d^4 x \sqrt{-g} \left[ \frac{1}{2}R^{\mu \nu}\bar{\nu}_R\gamma_\mu  D_\nu \nu_L + \frac{1}{4}\partial_\mu R \bar{\nu}_R\gamma^\mu \nu_L + \bar{\nu}_R D^2  \slashed{D} \nu_L \right] \\
\int d^4 x  \sqrt{-g} \left(\bar{\nu}_R\slashed{D} D^2 \nu_L\right)  &=\int d^4 x \sqrt{-g} \left[ \frac{1}{4}\partial_\mu R \bar{\nu}_R\gamma^\mu \nu_L + \bar{\nu}_R D^2  \slashed{D} \nu_L\right] \label{dd2}
\end{align}
In summary then, the list of linearly independent bilinears reduces to
\begin{equation}\label{basis}
\partial_\mu R \bar{\nu}_R \gamma^\mu \nu_L, \quad R \bar{\nu}_R\slashed{D}\nu_L,  \quad R_{\mu \nu}\bar{\nu}_R \gamma^\mu D^\nu\nu_L, \quad \bar{\nu}_R i D^2 \slashed{D}\nu_L,
\end{equation}
and so the most general form of the effective action is
\begin{align}
\Gamma = \int d^4 x \sqrt{-g(x)} \Bigg[& \bar{\nu}_R i \slashed{D}\nu_L + \alpha_1  \partial_\mu R \bar{\nu}_R \gamma^\mu \nu_L 
 + \alpha_2 i R \bar{\nu}_R\slashed{D}\nu_L + \alpha_ 3 \bar{\nu}_R R_{\mu \nu}\gamma^\mu i D^\nu\nu_L + \alpha_4 \bar{\nu}_R i D^2 \slashed{D}\nu_L \Bigg],
\end{align}
where the $\alpha_i$ can in general be complex. In order to respect the hermiticity of the full electroweak theory, we must demand that the effective action is hermitian and impose
\begin{equation}\label{hermit}
\Gamma^\dagger=\Gamma.
\end{equation}
At this point our analysis improves on that of Ohkuwa
\cite{Ohkuwa:1980jx} who did not impose the requirement of hermiticity
in constructing the effective action. 
The first operator in (\ref{basis}) is hermitian but the remaining operators are not. They have the following hermiticity properties:
\begin{align}
\int d^4 x  \sqrt{-g} \left( R\bar{\nu}_R i \slashed{D} \nu_L\right)^\dagger & =  \int d^4 x  \sqrt{-g}\left[ R \bar{\nu}_R i \slashed{D} \nu_L +   i\partial_\mu R \bar{\nu}_R \gamma^\mu \nu_L\right],  \\
\int d^4 x  \sqrt{-g} \left( R_{\mu \nu} \bar{\nu}_R \gamma^\mu  i D^\nu  \nu_L\right)^\dagger & =  \int d^4 x  \sqrt{-g} \left[ R_{\mu \nu} \bar{\nu}_R\gamma^\mu  i D^\nu  \nu_L + \frac{1}{2}i \partial_\mu R \bar{\nu}_R\gamma^\mu \nu_L\right],  \\
\int d^4 x  \sqrt{-g}\left( \bar{\nu}_R i D^2 \slashed{D}\nu_L \right)^\dagger &=\int d^4 x  \sqrt{-g} \left[\bar{\nu}_R i D^2 \slashed{D}\nu_L + \frac{1}{4}i \partial_\mu R \bar{\nu}_R\gamma^\mu \nu_L \right],
\end{align}
We now use these properties and impose the condition (\ref{hermit}) to get relations among the $\alpha_i$. We find that all the coefficients are all real, with the exception of $\alpha_1$, whose imaginary part must satisfy
\begin{equation}
\mbox{Im}(\alpha_1) = \frac{1}{2}\alpha_2 + \frac{1}{4}\alpha_3+ \frac{1}{8}\alpha_4.
\end{equation}
Hence with a suitable redefinition of the effective coefficients the most general form of the effective action is
\begin{align}
\Gamma = \int d^n x  \sqrt{-g(x)}   
\Bigg[   \bar{\nu}_Ri \slashed{D}\nu_L  &+ ia\bar{\nu}_R\left(2R_{\mu
    \nu}\gamma^\mu D^\nu   
+ \frac{1}{2}\partial_\mu R \gamma^\mu \right) \nu_L \nonumber\\
&+ b \partial_\mu R  \bar{\nu}_R \gamma^\mu \nu_L \nonumber\\
&+ ic \bar{\nu}_R\left( 2R \slashed{D}  +  \partial_\mu R \gamma^\mu
\right) \nu_L  \nonumber\\
& + id \bar{\nu}_R\left( 2 D^2\slashed{D}+ \frac{1}{4}\partial_\mu R \gamma^\mu \right)\nu_L  \Bigg]. \label{eff}
\end{align} 
The operators
\begin{align} 
S_a =& \int d^n x  \sqrt{-g(x)}~ i \bar{\nu}_R\left[2R_{\mu \nu}\gamma^\mu D^\nu  + \frac{1}{2}\partial_\mu R \gamma^\mu \right] \nu_L \\
S_b =& \int d^n x  \sqrt{-g(x)}~ \partial_\mu R  \bar{\nu}_R \gamma^\mu \nu_L \\ 
S_c =&\int d^n x  \sqrt{-g(x)}~ i \bar{\nu}_R \left[ 2R \slashed{D}  +  \partial_\mu R \gamma^\mu \right]\nu_L \\
S_d =& \int d^n x  \sqrt{-g(x)}~ i \bar{\nu}_R\left[ 2 D^2\slashed{D}+ \frac{1}{4}\partial_\mu R \gamma^\mu \right]\nu_L
\end{align}
which appear in $\Gamma$, therefore form a complete set of linearly
independent hermitian operators 
up to third order in derivatives.\footnote{Notice that instead of
  $S_a$, $S_c$ and $S_d$, we could alternatively have used the
  following basis of independently hermitian operators:
$i R_{\mu\nu} \bar{\nu}_R \gamma^\mu \overleftrightarrow{D}^\nu
\nu_L$,~
$i R \bar{\nu}_R \overleftrightarrow{\slashed{D}} \nu_L$ and 
$i \left(D_\mu \bar{\nu}_R\right) \overleftrightarrow{\slashed{D}} 
D^\mu \nu_L$,~
but these are less convenient for the subsequent application to the
matching conditions and equations of motion.}

\subsection{Discrete Symmetries}
Since the discrete spacetime symmetries P and T single out particular directions in spacetime, we assume the existence of a vector basis with a timelike vector $e^{(0)}$ and spacelike vectors $e^{(i)}$ which define spatial and temporal directions at each point $x$ in the manifold. We can then define P and T transformations locally at each point $x$ by:
\begin{align}\label{P}
P:& \qquad \mathcal{P} e^{(a)}(x) \mathcal{P}^{-1}= (-)^{a} e^{(a)}(x),  \qquad \mathcal{P}\nu(x)\mathcal{P}^{-1} = \gamma^0 \nu(x),\\
T:& \qquad \mathcal{T} e^{(a)}(x) \mathcal{T}^{-1}= -(-)^{a} e^{(a)}(x),  \qquad \mathcal{T} \nu(x) \mathcal{T}^{-1} = B \nu(x),\\
C:& \qquad \mathcal{C} \nu (x) \mathcal{C}^{-1}= C \left( \bar{\nu}\right)^T(x),
\end{align}
where the notion $(-)^a$ is a shorthand defined by
\begin{equation}
(-)^0 = 1, \qquad (-)^{i}= -1, \qquad i=1,2,3.
\end{equation}
The matrices satisfy $B^\dagger \gamma^{a*} B =(-1)^a \gamma^a$ and
$B^\dagger B=1$, 
$C \gamma^\mu C^{-1} = - \gamma^\mu$ and $C^T = -C$, and $\mathcal{T}$ complex conjugates any complex
numbers. This has the consequence that tensor quantities $T^{a_1 \ldots a_m} $ transform as 
\begin{equation}\label{tensorex}
T^{a_1 \ldots a_m}(x) \mapsto \left(\pm1\right)^{m} (-)^{a_1}\cdots (-)^{a_m}T^{a_1 \ldots a_m}(x)
\end{equation}
where the plus and minus sign correspond to P and T respectively. In particular, since we want $D_\mu \psi$ to transform like $\partial_\mu \psi$ under $P$ and $T$, it is easy to check that identifying the vector basis above with the vierbein ensures the connection part
\begin{equation}
\Gamma_c ~\psi  = -\frac{i}{4}\omega_c^{ab}\sigma_{ab}\psi,
\end{equation}
(and hence $D_\mu \psi$) transforms like $\partial_\mu \psi$ under P and T.

Notice that the arguments of the operators do not transoform as $(t,\vec{x}) \rightarrow (t , -\vec{x}) $, under P etc.~as they do in flat space. The reason is that in flat space, the object $x^\mu$ is playing the role of a vector (rather than a coordinate) so that $(t,\vec{x}) \rightarrow (t , -\vec{x}) $ should be thought of as an action on the tangent plane, rather than on coordinates. With an understanding of this subtlety, the generalisation to curved space is immediate, and one sees that P and T are only well-defined as actions on vectors in the tangent plane. For example, in the case of a scalar field, the object $\partial^\mu \phi (x)$ should be thought of as a vector in the tangent plane at $x$, which transforms according to (\ref{tensorex}).

We summarise the C, P, CP and CPT properties of the effective
operators in a table below. A full derivation of these results can be found in appendix B:
\bgroup
\def\arraystretch{1.5}
\begin{center}
\begin{tabular}{l*{3}{c}r}
            & $ \int d^4 x \sqrt{-g}~ \bar{\nu}_R i \slashed{D} \nu_L $
            &   $S_{a,c,d}$ &  $S_b$ \\
\hline
P       & $ L \leftrightarrow R $  & $ L \leftrightarrow R $ & $L \leftrightarrow R$ \\
T       &+1                              & $ +1$         & $-1$  \\
C        &$ L \leftrightarrow R $  & $L \leftrightarrow R$  &  $-\left(L \leftrightarrow R \right)$   \\
CP      & $+1 $                        & $+1$  &  $-1$ \\
CPT    &$+1 $                         & $+1$ &  $+1$ \\
\end{tabular}
\end{center}
\egroup
\noindent We see that with the exception of 
$L_b = \partial_\mu R \bar{\nu}_R \gamma^\mu \nu_L$ 
(which importantly is CP odd) all the operators respect the CP, T and CPT symmetries of the tree-level EW Lagrangian. It is thus of great interest to investigate the possibility that the CP violating operator $L_b$ is radiatively generated by quantum loop effects. This is particularly pertinent in light of the suggestion by Davoudiasl \textit{et al.} \cite{Davoudiasl:2004gf} that effectively generated C and CP-violating operators such as
\begin{equation}\label{dav}
{L}_{int} \sim \partial_\mu R \bar{\psi}\gamma^\mu \psi,
\end{equation}
give a chemical potential of the form $\mu \sim \dot{R}$ resulting in a gravitationally induced lepton or baryon asymmetry.

\subsection{Matching}

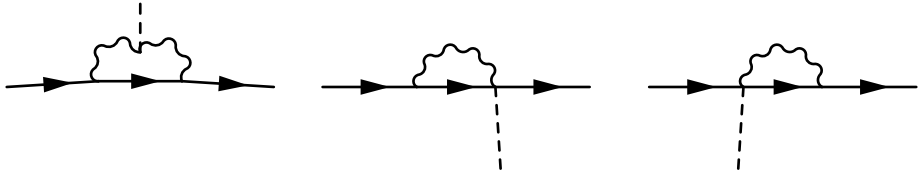
\begin{figure}[ht!]
    \centering
 \subfloat{
        \begin{fmffile}{graviton3}
\begin{fmfgraph*}(100,70)
\fmfleft{i}
\fmfright{o}
\fmftop{g}
\fmflabel{$p$}{i}
\fmflabel{$p'$}{o}
\fmflabel{$q$}{g}
\fmf{fermion,tension=2}{i,v1}
\fmf{fermion,tension=2}{v1,v2}
\fmf{fermion,tension=2}{v2,o}
\fmf{photon,left,tension=0.4}{v1,v3}
\fmf{photon,right,tension=0.4}{v2,v3}
\fmf{dashes,tension=0.4}{v3,g}
\end{fmfgraph*}
\end{fmffile}
    }
    \subfloat{
        \begin{fmffile}{graviton1}
\begin{fmfgraph}(100,70)
\fmfleft{i}
\fmfright{o}
\fmfbottom{g1,g2,g3,g4}
\fmf{fermion,tension=2}{i,v1}
\fmf{fermion,tension=2}{v1,v2}
\fmf{fermion,tension=2}{v2,o}
\fmf{photon,left,tension=0.4}{v1,v2}
\fmf{dashes,tension=0}{v2,g3}
\end{fmfgraph}
\end{fmffile}
    }
   \subfloat{
        \begin{fmffile}{graviton2}
\begin{fmfgraph}(100,70)
\fmfleft{i}
\fmfright{o}
\fmfbottom{g1,g2,g3,g4}
\fmf{fermion,tension=2}{i,v1}
\fmf{fermion,tension=2}{v1,v2}
\fmf{fermion,tension=2}{v2,o}
\fmf{photon,left,tension=0.4}{v1,v2}
\fmf{dashes,tension=0}{v1,g2}
\end{fmfgraph}
\end{fmffile}
    }
\caption{Graviton diagrams. The dashed line denotes momentum space contribution from classical graviton. The internal lines denote either $W$ and $e$ or  $Z$ and $\nu$.}
\label{grav}
\end{figure}

We now calculate the coefficients of the curvature terms in the effective action by matching
with explicit weak-field Feynman diagram calculations. 
Since the effective couplings are independent of the choice of
geometry, 
we are free to perform the matching on the most convenient
background, providing it is of sufficient generality to distinguish
the various terms in the action. 
The matching is greatly simplified by choosing a conformally flat metric
\begin{equation}
g_{\mu \nu} = K\eta_{\mu \nu},
\end{equation}
and conformally rescaled fields
\begin{equation}
e = K^{-(n-1)/4}\tilde{e}, \quad Z_{\mu} = K^{-(n-4)/4}\tilde{Z}_\mu,
\end{equation}
and similarly for the other fields. In terms of the conformally rescaled fields, the Lagrangian becomes
\begin{align}
\mathcal{L}_{EW} = & \mathcal{L}_{Maxwell} + \eta^{\mu \nu} K \left( M_W^2 \tilde{W}_\mu^-\tilde{W}^+_\nu + \frac{1}{2} M_Z^2 \tilde{Z}_\mu \tilde{Z}_\nu \right) \nonumber\\ 
& + \bar{\tilde{e}} i \slashed{\partial}\tilde{e} + \bar{\tilde{\nu}}_R i \gamma \slashed{\partial} \tilde{\nu}_L\nonumber\\ 
& + \frac{g}{\sqrt{2}} K^{-(n-4)/4}\left( \bar{\tilde{\nu}}_R \slashed{\tilde{W}}^+  \tilde{e}_L + \mbox{h.c} \right)	  -\frac{g}{2 \cos \theta_W} K^{-(n-4)/4} \bar{\tilde{\nu}}_R \slashed{\tilde{Z}}  \tilde{\nu}_L, 
\end{align}
where
\begin{align}
\mathcal{L}_{Maxwell} =\eta^{\mu \rho} \eta^{\nu \sigma} \Bigg[ & -\frac{1}{4} \tilde{Z}_{\mu \nu}\tilde{Z}_{\rho \sigma} \nonumber \\
&+\frac{(n-4)}{8}K^{-1}\left(\partial_\mu K \cdot \tilde{Z}_\nu - \partial_\nu K \cdot \tilde{Z}_\mu \right) \tilde{Z}_{\rho \sigma}\nonumber \\
& - \frac{(n-4)^2}{64}K^{-2}\left(\partial_\mu K \cdot \tilde{Z}_\nu - \partial_\nu K \cdot \tilde{Z}_\mu \right) \left( \partial_\rho K \cdot \tilde{Z}_\sigma - \partial_\sigma K \cdot \tilde{Z}_\rho \right) \nonumber \\ 
&+ (\mbox{similar term for $\tilde{W}^\pm$}) \Bigg]  \label{Max},
\end{align}
with $\tilde{Z}_{\mu \nu} = \partial_\mu \tilde{Z}_\nu - \partial_\mu \tilde{Z}_\nu$.
We match by considering 
\begin{equation}
K = 1+ h,
\end{equation}
and demanding that the full and effective theories give the same
amplitudes at linear order in the classical graviton $h$. There are 3 diagrams linear in $h$ shown in figure \ref{grav}

\pagebreak
\noindent We use $p^\mu$ and $p'^\mu$ to denote the incoming and outgoing neutrino momenta, and $q^\mu = p^\mu - p'^\mu$ to label the graviton momentum. The first of these gives an amplitude of the form\footnote{The general form of the fermion vertex is $\mathcal{L}_{int} = g_0K^{-(n-4)/4} \bar{\psi} \slashed{A}(1-\gamma_5)/2  \varphi$ for spinors $\psi$ and $\varphi$, and we insert appropriate couplings for $g_0$ later.}
\begin{align}
i \mathcal{M}_1  = & \left(i{M^2}\right)(-ig_0)^2 \int \frac{d^n k
}{(2\pi)^n} \frac{i}{(p-k)^2-M^2}
\frac{i \gamma_\mu \slashed{k}\gamma^\mu}{k^2}\frac{i}{(p-k-q)^2-M^2} \nonumber\\
\nonumber\\
 = &  \frac{ i g_0^2  }{ (4\pi)^2 M^2} \left(
\frac{1}{2}\slashed{q}M^2
- \slashed{p}M^2  
+ \frac{1}{6}q^2 \slashed{q}
+ \frac{1}{6}p^2\slashed{q}
- \frac{2}{9} p\cdot q \slashed{q}
- \frac{5}{18}q^2 \slashed{p} 
+ \frac{1}{3} p\cdot q \slashed{p}  
- \frac{1}{3} p^2 \slashed{p}  \right)  \nonumber \\
& + \mathcal{O}\left(\frac{p^5}{M^4} \right)
\end{align}
The second diagram gives
\begin{equation}
i \mathcal{M}_2 = \left(ie\frac{(n-4)}{4}h\right)(-ie) \int \frac{d^n
  k}{(2\pi)^n}\frac{i}{(k-p)^2 -M^2}  \frac{i\gamma_\mu
  \slashed{k}\gamma^\mu}{k^2} 
~~\stackrel{n\rightarrow 4}{=} ~~ i \frac{g_0^2}{2 (4\pi)^2} \slashed{p},
\end{equation}
where the $(n-4)$ conformal factor from the graviton vertex combines with the
UV divergence of the self energy diagram to produce a finite result.
The third diagram is the same as this, but with $p \rightarrow p-q$:
\begin{equation}
i \mathcal{M}_3 =  i \frac{g^2_0}{2 (4\pi)^2} \left( \slashed{p} - \slashed{q} \right). 
\end{equation}
When we sum the three diagrams, the terms linear in momentum cancel so
there is no finite field renormalization of the free Dirac operator in
the effective Lagrangian.
\begin{equation}
\mathcal{M}_{tot} = \frac{ g_0^2  }{ 3 (4\pi)^2 M^2} \left(
\frac{1}{2}q^2 \slashed{q}
+ \frac{1}{2}p^2\slashed{q}
- \frac{2}{3} p\cdot q \slashed{q}
- \frac{5}{6}q^2 \slashed{p} 
+  p\cdot q \slashed{p}  
- p^2 \slashed{p}  \right)
\end{equation}
We now insert the appropriate couplings and count this amplitude twice
for the different particle species in the loop, 
one with $g_0=g/\sqrt{2}$, $M=m_W$ and the other with $g_0=g/(2 \cos
\theta_W)$, $M=m_Z$. 
Using the result $m_W^2 = \cos^2 \theta_W m_Z ^2$ the full amplitude can be written in terms of the SU(2) coupling $g$ and the $W$-boson mass as
\begin{equation} 
\label{UV}
 \mathcal{M}= \frac{1}{4}\frac{ g^2 }{ (4\pi)^2 m_W^2} \left[     \left(    -  p^2   +  p\cdot q   - \frac{5}{6}q^2  \right) \slashed{p} + \left(   \frac{1}{2}p^2 - \frac{2}{3}p\cdot q + \frac{1}{2}q^2\right)  \slashed{q}   \right].
\end{equation}

\subsection*{Poles and The Maxwell Term}
One might worry that the vertex proportional to $(n-4)$ in the Maxwell term (\ref{Max}) gives a finite contribution to the amplitude for the first diagram in figure \ref{grav} due to poles in the loop integral. However, we now show this contribution vanishes. We prove this for the $Z$-boson, with the $W$ following in exactly the same way. The relevant interaction is
\begin{align}
\mathcal{L}_{int} & = \frac{(n-4)}{8}K^{-1}\left(\partial_\rho K \cdot \tilde{Z}_\lambda - \partial_\lambda K \cdot \tilde{Z}_\rho \right) \tilde{Z}^{\rho \lambda}\\
& = \frac{(n-4)}{4} \partial_\rho h \tilde{Z}_\lambda \left[   \partial^\rho \tilde{Z}^\lambda -  \partial^\lambda \tilde{Z}^\rho   \right],
\end{align}
The amplitude for $\mathcal{L}_{int}$ is given by
\begin{align}
i\mathcal{M}_{pole} = & \frac{(n-4)}{4}\left(- \frac{ig}{2 \cos \theta_W} \right)^2 \times \\
&\int \frac{d^nk}{(2\pi)^n}\Bigg[  \left[q \cdot (p-k-q) - q \cdot (p-k) \right] D_{\mu \lambda}(p-k)\tensor{D}{^\lambda _ \sigma}(p-k-q) \nonumber \\
&  \qquad - q^\rho (p-k-q)^\lambda D_{\mu \lambda}(p-k) D_{\rho \sigma}(p-k-q)\nonumber  \\
& \qquad +  q^\rho (p-k)^\lambda  D_{ \mu \rho}(p-k)  D_{\lambda \sigma}(p-k-q) \Bigg] \gamma^\mu \frac{i\slashed{k}}{k^2}\gamma^\sigma,
\end{align}
where $D_{\mu \nu}(p)$ is the $\tilde{Z}$ propagator:
\begin{equation}\label{prop}
D_{\mu \nu}(q) =\frac{i \eta_{\mu \nu}}{q^2 - m_Z^2}.
\end{equation} 
Inserting (\ref{prop}), the amplitude reduces to
\begin{align}
i\mathcal{M}_{pole} =  \frac{(n-4) }{16 \cos^2 \theta_W}g^2  \int\frac{d^n q}{(2 \pi)}  & \Bigg[  q_\mu (p-k)_\sigma -  q^2\eta_{\mu \sigma}  - q_\sigma  (p-k-q)_\mu  \Bigg] \nonumber \\
& \times \frac{i}{(p-k)^2 - m_Z^2} \gamma^\mu \frac{i\slashed{k}}{k^2}\gamma^\sigma \frac{i}{(p-k-q)^2 - m_Z^2} 
\end{align}
Only terms in the momentum integral which produce a pole in $(n-4)$ will contribute. These come from the highest powers of loop momenta. Retaining only those terms we get
\begin{align}
i\mathcal{M}_{pole} = &  \frac{(n-4) }{16 \cos^2 \theta_W}g^2 \int\frac{d^n q}{(2 \pi)}   \left[   k_\mu q_\sigma - k_\sigma q_\mu \right]  \frac{i}{(p-k)^2 - m_Z^2} \gamma^\mu \frac{i\slashed{k}}{k^2}\gamma^\sigma \frac{i}{(p-k-q)^2 - m_Z^2} \nonumber \\ 
&+ O(n-4),
\end{align}
but
\begin{equation}
 \left[   k_\mu q_\sigma - k_\sigma q_\mu \right]\gamma^\mu \slashed{k}\gamma^\sigma =0,
\end{equation}
and so the momentum integral vanishes, verifying our claim that there is no contribution in the $n \rightarrow 4$ limit from the Maxwell term.

\noindent One can in fact verify the general form of this amplitude (\ref{UV}) as follows. The amplitude for the above process is given by
\begin{equation}
\mathcal{M}(p,q) = \bra{\nu(p)} S \ket{h(q), \nu(p')}.
\end{equation}
It follows from the unitarity of the scattering matrix that
\begin{equation}\label{unit}
\mathcal{M}^* (p,q)= \bra{h(q), \nu(p')} S  \ket{\nu(p)} \equiv M(p-q,-q),
\end{equation}
where in the last equality we used the fact this amplitude describes an incoming neutrino of momentum $p'=p-q$ and incoming graviton of momentum $q$. If we write
\begin{equation}\label{form}
\mathcal{M}(p,q) = \left(\alpha p^2 + \beta p \cdot q + \gamma q^2  \right)\slashed{p} + \left(\delta p^2 + \varepsilon p \cdot q + \phi q^2  \right)\slashed{q}
\end{equation}
the relation $\mathcal{M}^* (p,q)=M(p-q,-q)$ in (\ref{unit}) implies the following relations amongst the coefficients in (\ref{form})
\begin{align}
\beta & = - \alpha, \label{h1}\\
\delta &= -\frac{1}{2}\alpha, \label{h2}\\
\phi & = \frac{1}{4}\alpha - \frac{1}{2} \gamma - \frac{1}{2}\varepsilon. \label{h3}
\end{align}
and it is easily checked that the corresponding values in (\ref{UV})
satisfy these relations. 

We now compute the contribution from the graviton vertex in the
effective action.

\begin{figure}[h]
\centering
\begin{fmffile}{effective}
\begin{fmfgraph}(70,55)
\fmfleft{i}
\fmfright{o}
\fmftop{g}
\fmf{fermion,tension=1}{i,v}
\fmf{fermion,tension=1}{v,o}
\fmf{dashes,tension=0}{v,g}
\fmfdot{v}
\end{fmfgraph}
\end{fmffile}
\caption{The effective graviton vertex.}
\end{figure}
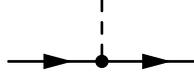
\noindent In order to do this we must expand the various neutrino-curvature operators to linear order in $h$ in terms of the tilde fields. Using the relation
\begin{equation}
R_{\mu \nu} = - \partial_\mu \partial_\nu h - \frac{1}{2}\eta_{\mu \nu}\partial^2 h + O(h^2),
\end{equation}
we find the following results
\begin{align}
 \bar{\nu}_R R_{\mu \nu}\gamma^\mu D^\nu \nu_L=& -  \bar{\tilde{\nu}}_R \left[   \partial_\mu \partial_\nu h  \gamma^\mu \partial^\nu + \frac{1}{2} (\partial^2 h) \slashed{\partial}\right]\tilde{\nu}_L, \nonumber\\
\partial_\mu R \bar{\nu}_R \gamma^\mu  \nu_L  =& - 3 \bar{\tilde{\nu}}_R \partial^2 \slashed{\partial}h \tilde{\nu}_L, \nonumber\\
 R \bar{\nu}_R \slashed{D}\nu_L:  =& - 3   \partial^2 h \bar{\tilde{\nu}}_R \gamma\cdot \partial  \tilde{\nu}_L, \nonumber\\
\sqrt{-g}\bar{\nu}_R D^2 \slashed{D}  \nu_L =& \Bigg \lbrace  \bar{\tilde{\nu}}_R\left[ - h \partial^2  +  \partial_\mu h \partial^\mu  - \frac{i}{2} \sigma_{ab} \partial^b h \partial^a \right] \slashed{\partial}\tilde{\nu}_L +  \bar{\tilde{\nu}}_R\partial^2 \left[-\frac{1}{2}h\slashed{\partial} + \frac{3}{4}\slashed{\partial}h \right] \tilde{\nu}_L \nonumber\\
&  + \frac{5}{4}h\bar{\tilde{\nu}}_R \partial^2 \slashed{\partial}\tilde{\nu}_L  - \frac{3}{4}\bar{\tilde{\nu}}_R\partial^2 \slashed{\partial} (h \tilde{\nu}_L) \Bigg \rbrace.
\end{align}
The final result for $L_d$ is more complicated than the others since
there is an $O(h)$ contribution from the spin connection. A careful derivation can be found in appendix A. 

Using these expressions we can use the effective action (\ref{eff}) to calculate the total momentum space contribution for the effective vertex as
\begin{equation}\label{effec}
\mathcal{M}_{eff} =  \left(2d p^2 - 2 d p\cdot q  +\left[a + 6 c + \frac{5}{2}d \right]q^2 \right) \slashed{p} +  \left(   - d p^2  + 2 a p\cdot q + \left[ 3ib - \frac{3}{2}a-3c -\frac{3}{4}d \right]q^2\right) \slashed{q} 
\end{equation}
The full amplitude has no imaginary part and so $b=0$. The remaining
effective coefficients are given by solving 6 equations in the
remaining 3 unknowns $a,c,d$ given by comparing (\ref{effec}) with
(\ref{UV}). In fact, the hermiticity constraints mean that the number of independent equations is actually 3
rather than 6, due to the inter-relations  (\ref{h1}) -- (\ref{h3}) amongst the effective couplings. The equations are solved by
\begin{equation}
a = -\frac{1}{12}\frac{g^2}{(4\pi)^2 m_W^2}, \qquad c= \frac{1}{32}\frac{g^2}{(4\pi)^2 m_W^2}, \qquad d = -\frac{1}{8}\frac{g^2}{(4\pi)^2 m_W^2},
\end{equation}
thus completing our calculation of the effective action to $O(g^2)$. In so doing we have verified the validity of the form of the effective action, and explicitly calculated the strength of these strong equivalence violating curvature couplings. In the following section we discuss the physical interpretation of this violation and look at its implications for the nature of neutrino propagation.

\subsection*{Discrete Symmetries Revisited}

The lack of an imaginary part of the loop amplitude means that $L_b$
(which is CP odd) finds no support. As a result, the only remaining
operators are $L_a,L_c$ and $L_d$, which respect the tree-level
symmetries CP and CPT of the tree-level action.
The final form of $\Gamma$ is given by:
\begin{align}
\Gamma = \int d^4 x  \sqrt{-g(x)}   
\Bigg[  & \bar{\nu}_Li \slashed{D}\nu_L
-\frac{1}{12}\frac{g^2}{(4\pi)^2 m_W^2}\bar{\nu}_R i \left(2R_{\mu
    \nu}\gamma^\mu D^\nu   + \frac{1}{2}\partial_\mu R \gamma^\mu
\right) \nu_L \nonumber\\
&+ \frac{1}{32}\frac{g^2}{(4\pi)^2 m_W^2}\bar{\nu}_Ri \left( 2R
  \slashed{D}  +  \partial_\mu R \gamma^\mu \right) \nu_L \nonumber\\
& -\frac{1}{8}\frac{g^2}{(4\pi)^2 m_W^2} \bar{\nu}_R i \left( 2 D^2\slashed{D}+ \frac{1}{4}\partial_\mu R \gamma^\mu \right)\nu_L  \Bigg]. 
\end{align} 

We infer from this that even though it is possible to construct
hermitian operators which would break them, the discrete symmetries 
C and CP are {\it not} anomalously broken by the strong equivalence
violating interactions generated by quantum loops. 
It therefore appears that to induce the C and CP-violating curvature 
couplings required for gravitational leptogenesis,
there must already be a source of symmetry violation in the original
tree-level theory.

Moreover, the analysis above shows how it may be misleading to
consider a single symmetry-violating curvature interaction in
isolation, since only when the complete set of hermitian operators is
considered is it possible to determine the coefficients in the
effective action and assess whether the discrete symmetries are 
conserved or broken.

\section{Dispersion Relations, Neutrino Propagation and Leptogenesis Models}

In this section we return to the analysis of dispersion relations
introduced in section 2 and show how the eikonal approximation is
modified in the presence of various strong equivalence violating
curvature interactions. We discuss both those generated in the
effective action arising from quantum loop effects and the CP-violating
interactions introduced by hand in the leptogenesis model of ref.\cite{Davoudiasl:2004gf}.

\subsection{Gravitational effects on neutrino propagation}

We begin with a discussion of the propagation of massless neutrinos in the
standard model, based on the effective action above. We show how
the gravitational tidal forces encoded in the low-energy effective
action result in superluminal neutrino propagation.

First, we note that since 
\begin{equation}
\slashed{D}\nu_L = 0 + O(g^2), 
\end{equation}
the terms
\begin{equation}
d i D^2 \slashed{D} \nu_L, \qquad c R \slashed{D}\nu_L,
\end{equation}
only affect the equations of motion at $O(g^4)$. As a result, the $O(g^2)$ equation of motion is
\begin{equation}
\left[i\slashed{D} + 2 ai R_{\mu \nu} \gamma^\mu D^\nu  + i
  \tilde{b}\slashed{\partial} R  \right]\nu_L =0
\label{tildeb}
\end{equation}
where the factor $\tilde{b}$ can be determined from $\Gamma$ by
collecting like terms:
\begin{equation}
\tilde{b} = -\frac{1}{24}\frac{g^2}{(4\pi)^2 m_W^2}
\end{equation}
We now insert the eikonal ansatz
\begin{equation}
\nu_L =   \left( a^{(s)} -i \epsilon b^{(s)} -\epsilon^2 c^{(s)} +\ldots \right)u^{(s)} e^{-i \Theta/\epsilon},
\end{equation}
which gives
\begin{align}
O(1/\epsilon): \qquad & \left(g_{\mu \nu} + 2 a R_{\mu \nu}
\right)\gamma^\mu k^\nu a\cdot u =0  \\
O(1) : \qquad &\left(g_{\mu \nu} + 2a  R_{\mu \nu}  \right)\gamma^\mu D^\nu a\cdot u  
+ i \tilde{b} \gamma^\mu\partial_\mu R a\cdot u  
+ \left(g_{\mu \nu} + 2 a R_{\mu \nu}
\right)\gamma^\mu k^\nu b\cdot u =0,
\end{align}
where, as usual, the wave-vector is $k_\mu = \partial_\mu \Theta$.
The leading order term gives the quantum corrected dispersion
relation. To $O(g^2)$, we find
\begin{equation}
k^2 + 4 a R_{\mu \nu} k^\mu k^\nu  = 0,
\label{disp}
\end{equation}
which is elegantly rewritten in terms of an effective metric
$G_{\mu\nu} = g_{\mu\nu} + 4a R_{\mu\nu}$, since then we have simply
\begin{equation}
G_{\mu\nu} k^\mu k^\nu = 0 \ .
\end{equation}
This `bimetric' interpretation has previously been developed in the
context of photon propagation in curved spacetime in, 
e.g.~\cite{Drummond:1979pp,Shore:2003jx}.

To deduce the evolution of the amplitude, it is more convenient to use
the method described in section 2 of squaring the Dirac equation and
analysing the resulting wave equation. Including the curvature terms,
this gives:
\begin{align}
O(1/\epsilon^2) : &&   \left(G_{\mu\nu}k^\mu k^\nu\right) ~ a\cdot u =0 \\
O(1/\epsilon): && \left[G_{\mu\nu} k^\mu D^\nu  +
  \frac{1}{2}D^\mu(G_{\mu\nu}k^\nu) 
  -a \frac{i}{4} k^\mu D^\nu R_{\mu\nu\lambda\rho}\sigma^{\lambda\rho}
  +\tilde{b}k^\mu D_\mu R    \right] a\cdot u ~~~~\nonumber\\
&& -\left(G_{\mu\nu} k^\mu
  k^\nu\right)~b\cdot u=0 
\label{neutrino}
\end{align}
where the $b\cdot u$  term in
(\ref{neutrino}) vanishes when the dispersion relation equation
is applied (see (\ref{curv})). The $O(1)$ term is readily derived and describes
higher order variations of the amplitude. 

The first two terms in (\ref{neutrino}) are a straightforward
generalisation of the earlier result that the amplitude evolves along
the trajectory according to the expansion of the geodesic congruence.
Apart from the occurrence of the effective metric $G_{\mu\nu}$, the
significant difference here is that due to the Ricci curvature term in
the equation of motion, the amplitude evolution now
involves the uncontracted tensor $\Omega_{\mu\nu} = D_\mu k_\nu$, 
so that {\it both} the expansion and shear are involved. 
There are additional contributions 
from the Ricci term depending on the spin-curvature interaction, 
and also from the variation of the Ricci scalar along the trajectory.

In this case, therefore, the eikonal approach gives a clear solution
describing a quasi-plane wave with both an amplitude modulation over
a length scale $L \sim 1/\sqrt{R} \gg \lambda$ {\it and} a phase
modulation over a similar scale. The dispersion relation is therefore
modified as in (\ref{disp}) with a consequent change in the phase velocity.

It is interesting to evaluate this explicitly for a cosmological
spacetime. First, for any background satisfying the Einstein
equations, note that the dispersion relation at $O(g^2)$ can be
written in terms of the background matter energy-momentum tensor
as 
\begin{equation}
k^2 + 4a T_{\mu\nu} k^\mu k^\nu = 0 \ .
\end{equation}
It follows that for any background satisying the null energy condition
$T_{\mu\nu} k^\mu k^\nu > 0$, the wave-vector is timelike,
$k^2 > 0$. Notice however that this implies the phase velocity 
$v_{\rm ph} = k^{0}/|\textbf{k}|$ (with components defined in the
vierbein frame) 
is superluminal.\footnote{See
  ref.\cite{Shore:2003jx} for a discussion of how this corresponds
  to a momentum interpretation, where $p^2 < 0$ is spacelike
  and the particle velocity $v = |\textbf{p}|/p^0$ is superluminal
  and equal to the phase velocity $v_{\rm ph}$ above.}

Expanding the dispersion relation (\ref{disp}) in components, we find
in general
\begin{equation}
k^0 = \vert \textbf{k} \vert \left[ 1 - 2a \left( R_{00} 
+ 2 R_{0i}\frac{k^i}{\vert \textbf{k} \vert} 
+  R_{ij}\frac{k^i k^j}{\vert \textbf{k} \vert^2}  \right) \right]\ .
\end{equation}
So in a FRW universe with energy density $\rho$ and pressure $p$
we find, restoring all factors, that the neutrino phase velocity is
\begin{equation}
v_{\rm ph}/c = 1 + \frac{2}{3}\frac{g_1^2}{(4\pi)^2 m_W^2}  
\frac{\hbar}{c^3 } \frac{8 \pi G}{c^4} \left( p + c^2 \rho \right) \ .
\end{equation}
This confirms the result first obtained by Ohkuwa \cite{Ohkuwa:1980jx}
that the neutrino velocity derived from the effective Lagrangian
(\ref{eff}) is superluminal. 

Note, however, that although we have reproduced the same 
result as Ohkuwa, it was far from obvious that this would be the case 
when the matching was imposed with the complete form of the 
effective action with {\it all} the hermitian operators included,
since in the original work \cite{Ohkuwa:1980jx} only the coefficient of the 
$L_a$ operator was considered. Moreover, by carrying out a fully
covariant eikonal analysis of the propagation equation for neutrinos,
we were able to determine not just the quantum correction to the
dispersion relation but also the evolution of the amplitude along the
neutrino trajectory, revealing its dependence on the operator
$L_{\tilde{b}}$.

Notice also that since the dispersion relation above 
is derived from a low-energy effective
action (recall that the Lagrangian (\ref{eff}) is a derivative
expansion, valid for energies below the $W$ mass), this gives the 
phase velocity in the low-frequency limit.
A full calculation, especially determining the high-frequency limit of
the phase velocity as required for discussions of causality, would
require the generalisation to spinors of the formalism developed in
refs.\cite{Shore:2007um,Hollowood:2007ku,
Hollowood:2008kq} to study the full frequency-dependence of the refractive 
index for photon propagation in curved spacetimes.

\subsection{CP-violating interactions and leptogenesis}

In our construction of the effective Lagrangian \ref{eff}, we found
that the CP-violating coupling $L_b = \partial_\mu R \bar\psi\gamma^\mu \psi$ 
did not appear if we started from a CP-conserving fundamental theory.
However, by including CP-violating couplings, {\it e.g.}~from neutrino
flavour mixing, in the original theory we could easily induce such terms
in $L_{\rm eff}$.  As explained earlier, this term is of particular
interest in gravitational leptogenesis. In a background with a
time-dependent Ricci scalar, a coupling $\frac{1}{M^2}\partial_\mu R
J^\mu$, where $J^\mu$ is a lepton number current, resembles a
chemical potential $\mu \sim \dot{R}$ term which in a conventional
flat space thermodynamic analysis would produce a matter-antimatter
asymmetry. This model, where $L_b$ is simply added by hand to the tree-level
action with $M$ being an arbitrary mass scale, has been studied in
refs.\cite{Davoudiasl:2004gf}.  

It is therefore interesting to examine the propagation equations for 
fermions in a theory with a term of type $L_b$ in the Dirac Lagrangian 
and to see how the CP violation and any matter-antimatter asymmetry
is manifested in the dispersion relation.

In this case, therefore, we start from the Lagrangian
\begin{equation}
\mathcal{L} = \bar\psi (i\slashed{D} - m + b \slashed{\partial}R)\psi = 0 \ ,
\label{Diracb}
\end{equation}
where in general $b$ is an arbitrary coupling of dimension $1/M^2$.
Note the all-important factor of $i$ difference from the $\tilde{b}$
term in the case above (compare (\ref{tildeb})).

As we now show, the appropriate way to solve the propagation equations
here is to generalise the conventional eikonal ansatz to
\begin{equation}
\psi(x) = \left(a^{(s)}(x) -i b^{(s)}(x)\epsilon + \ldots \right) u^{(s)}(x) 
e^{-\frac{i}{\epsilon}\left(\Theta(x) + \epsilon \alpha(x) + \ldots\right) }\ ,
\label{eikc}
\end{equation}
where we also allow the phase itself to have corrections which are
sub-leading in the counting parameter $\epsilon$.\footnote{Note 
this is quite different from the case above where the entire phase was
of order $1/\epsilon$, although $\Theta$ itself was expanded in
the subsidiary perturbative parameter $g^2 R/m_W^2$.}
Substituting this ansatz into the Dirac equation (\ref{Diracb}) and
setting $k_\mu = \partial_\mu \Theta$ as usual, we find
\begin{equation}
\frac{1}{\epsilon}(\slashed{k} - m)a.u + (\slashed{D}\alpha + b
  \slashed{R})a.u
+ i \left(\slashed{D} a.u  - (\slashed{k} - m) b.u\right)= 0 \ .
\end{equation}
The natural solution is to take 
\begin{equation}
{\rm det}(\slashed{k} - m) = 0
\end{equation}
leading to 
\begin{equation}
k^2 - m^2 = 0 \ ,
\label{dispk}
\end{equation}
together with $\alpha = -bR$, which removes the real part of the $O(1)$ term.
That is, we absorb the whole curvature
correction into the {\it phase}. In fact, this is immediately apparent from the 
Lagrangian, since we can remove the $L_b$ term by a change of variable
$\psi \rightarrow e^{ibR}\psi$. In turn, this simply reflects the fact 
that $L_b$ involves the fermion number current, corresponding to the
global symmetry $\psi \rightarrow e^{i\theta}\psi$.

To find the amplitude variation, it is most convenient to use the
squared Dirac equation. At $O(1/\epsilon^2)$ this reproduces the 
dispersion relation above while at $O(1/\epsilon)$ we have simply
\begin{equation}
\left(k\cdot D + \frac{1}{2}D\cdot k\right) a\cdot u = 0 \ .
\label{kaevol}
\end{equation}
The full solution is therefore
\begin{equation}
\psi(x) = a^{(s)} u^{(s)} e^{-i(\Theta(x) - b R(x))} \ .
\end{equation}
This is to be interpreted as a {\it phase modulation} of the
plane-wave solution of the free Dirac equation. It is still important
that the scale (in space or time) over which the frequency or
wavelength changes, which is set by the curvature, is much less
than the fundamental frequency so that we are still considering 
quasi-plane waves admitting an approximate particle interpretation.
The wave-vector $K_\mu$ for the quasi-plane wave, including the
corrections, is defined as the derivative of the entire phase, {\it i.e.}
$K_\mu = k_\mu - b \partial_\mu R$, and satisfies the modified
dispersion relation (from \ref{dispk})
\begin{equation}
(K_\mu + b\partial_\mu R)^2 - m^2 = 0 \ .
\end{equation}
At leading order, the amplitude satisfies the evolution equation 
(\ref{kaevol}), so in terms of the true wave-vector $K^\mu$ it propagates 
as usual according to the expansion scalar of the congruence
defined as the integral curves of $K^\mu + b \partial^\mu R$.

Unlike the previous cases we have considered, where it was sufficient
simply to look at the particle solutions, because we are dealing here
with a $C$ odd correction term in the Lagrangian we find a different
dispersion relation for the antiparticles. It is clear that the
antiparticle solution (with the spinor $v^{(s)}$) simply involves
reversing the sign of the curvature term in the phase and in the 
dispersion relation, so that the relations for particles/anti-particles are
\begin{equation}
(K_\mu \pm b\partial_\mu R)^2 - m^2 = 0,
\end{equation}
respectively.

This different phase modulation in the quasi-plane waves representing 
the particle and antiparticle solutions opens the door to using the
Lagrangian (\ref{Diracb}) as a source of matter-antimatter asymmetry in
realistic models of leptogenesis. As seen above, it is clearly closely
related to the original motivation for introducing the $L_b$
correction as an effective chemical potential for lepton number.

\section{Discussion}

In this paper, we have studied a number of fundamental theoretical
issues related to gravitational leptogenesis. In particular, we have
investigated whether the C, CP and CPT-violating operators necessary
to satisfy (or circumvent) the Sakharov conditions may be generated at
the quantum loop level through the mechanism of radiatively induced
strong equivalence principle breaking. 

The effective action for standard model neutrinos in curved spacetime
was constructed by careful matching of perturbative Feynman amplitudes in a 
weak background field to operators in an effective Lagrangian,
emphasising the need to consider a complete set of hermitian operators
to ensure consistency. The first aim was to look for any sign of 
``anomalous symmetry breaking'', {\it i.e.}~whether any operators were
induced at the quantum level which did not respect the discrete
symmetries of the original classical Lagrangian. In fact, we found no
evidence for this -- in particular, CP violating operators of the form
$\partial_\mu R \bar\psi \gamma^\mu \psi$, as required in the
leptogenesis model of refs.\cite{Davoudiasl:2004gf}, were shown not to be
generated in a theory with a CP conserving tree-level Lagrangian. 
It appears that if such interactions are to arise in an effective
Lagrangian, the underlying theory must already contain the seeds of
symmetry violation, {\it e.g.} in the form of explicit CP-violating 
coupling constants.

Radiatively-induced strong equivalence breaking interactions may
nevertheless be important for leptogenesis through what we termed
``environmental symmetry breaking''.  In a fixed background, the
curvature acts as a, possibly space or time-dependent, coupling to a
fermion bilinear operator which need not share the symmetry of the
combined term in the effective Lagrangian. For example, while the full
operator $\partial_\mu R \bar\psi \gamma^\mu \psi$ respects CPT, 
the fermion bilinear $\bar\psi \gamma^0 \psi$ itself is both CP and
CPT odd, so in a spatially homogeneous and isotropic background 
with a time-varying Ricci curvature $\dot R \neq 0$, effective 
CPT-varying physical effects with matter-antimatter asymmetry will arise.

The way these curvature interactions produce matter-antimatter
asymmetry is usually presented either, as in ref.\cite{Davoudiasl:2004gf},
by identifying the curvature as analogous to a chemical potential
for lepton number in conventional flat space thermodynamics, 
or by inferring a splitting in energy levels for particles and
antiparticles from a curvature-modified dispersion relation.

This motivated us to perform a detailed analysis of dispersion
relations for fermion theories in the presence of strong equivalence
breaking curvature couplings, whether introduced at tree level or
arising through quantum loops. We emphasised the importance of a 
fully-covariant description and the relation to physics in a local
inertial frame was highlighted; in particular, it was shown that in a 
Riemannian background the spin connection plays no role in the 
dispersion relation, contrary to some claims in the literature
\cite{Singh:2003sp,Mukhopadhyay:2005gb,Debnath:2005wk}. 
Our analysis was carried out in the framework
of the eikonal approximation, which clarifies how to incorporate the
hierarchy of scales characteristic of leptogenesis models.

Two models were considered in detail. In the case of neutrino
propagation in the standard model, it was possible to show that the
quantum loop effects do modify the leading eikonal term in the dispersion relation.
Using our complete effective Lagrangian, we were able to confirm an
earlier result due to Ohkuwa that the low-frequency phase
velocity for massless neutrinos is superluminal in a gravitational
background satisfying the null energy condition. Assuming causality is
respected, which requires the high-frequency limit of the phase
velocity to be $c$, this implies that the Kramers-Kronig relation is
violated for fermionic Green functions in curved spacetime, as has
previously been demonstrated for photon propagation in 
QED \cite{Shore:2007um,Hollowood:2007ku,
Hollowood:2008kq}.

Finally, we investigated a model where the curvature couples
directly to a lepton number current through the CP-violating interaction
$\partial_\mu R J^\mu$. Here, we saw how a generalisation of the
usual eikonal expansion shows that particle propagation is described
by phase-modulated quasi-plane waves in which the phase is modified 
by the Ricci scalar in the opposite way for particles and
antiparticles. In principle, therefore, this provides a mechanism for
gravitational leptogenesis.

\vskip1.5cm
\noindent{\bf Acknowledgements:}
\vskip0.5cm

We are grateful to the U.K.~Science and Technology Facilities Council
(STFC) for support under grants ST/J00040X/1, ST/L000369/1 and
ST/K50237.  We would like to thank G.~Aarts, J.~Ellis,
T.~Hollowood, N.~Mavromatos and S.~Sarkar  
for useful discussions.

\vskip1.5cm

\appendix

\section{Gravitational weak field expansion}

We derive the $O(h)$ expansion of the operator $ \sqrt{-g}\bar{\nu}_R D^2 \slashed{D}\nu_L$. Firstly we substitute for $\tilde{\nu}$ to get
\begin{equation}
\sqrt{-g}\bar{\nu}_R D^2 \slashed{D} \nu_L = \sqrt{-g}\bar{\tilde{\nu}}_R D ^2 \slashed{D} \tilde{\nu}_L - \frac{3}{4}h\bar{\tilde{\nu}}_R \partial^2 \slashed{\partial}\tilde{\nu}_L  - \frac{3}{4}\bar{\tilde{\nu}}_R\partial^2 \slashed{\partial} (h \tilde{\nu}_L)
\end{equation}
Next we expand $\sqrt{-g}= 1 + 2h$
\begin{equation}
\sqrt{-g}\bar{\nu}_R D^2 \slashed{D} \nu_L = \bar{\tilde{\nu}}_R D^2 \slashed{D} \tilde{\nu}_L + \frac{5}{4}h\bar{\tilde{\nu}}_R \partial^2 \slashed{\partial}\tilde{\nu}_L  - \frac{3}{4}\bar{\tilde{\nu}}_R \partial^2 \slashed{\partial} (h \tilde{\nu})_L
\end{equation}
Then we expand the first term as
\begin{equation}
\sqrt{-g} \bar{\tilde{\nu}}_R D^2 \slashed{D} \tilde{\nu}_L = \bar{\tilde{\nu}}_R \left[ (1- h) \eta^{\mu \nu}\partial_\mu \partial_\nu  +  \partial_\mu h \partial^\mu  - \frac{i}{2} \sigma_{ab} \partial^b h \partial^a \right] 
 \left[\eta^{a \nu}\gamma_a \partial_\nu -\frac{1}{2}h\gamma_a\partial^a + \frac{n-1}{4}\slashed{\partial}h\right]\tilde{\nu}_L
\end{equation}
Keeping only the linear $h$ term we get,
\begin{equation}
\sqrt{-g} \bar{\tilde{\nu}}_R D^2 \slashed{D} \tilde{\nu}_L =  \bar{\tilde{\nu}}_R \left[ - h \eta^{\mu \nu}\partial_\mu \partial_\nu  +  \partial_\mu h \partial^\mu  - \frac{i}{2} \sigma_{ab} \partial^b h \partial^a \right] \slashed{\partial}\tilde{\nu}_L +  \bar{\tilde{\nu}}_R \partial^2 \left[-\frac{1}{2}h\gamma_a\partial^a + \frac{n-1}{4}\slashed{\partial}h \right] \tilde{\nu}_L
\end{equation}
Putting this together we find
\begin{align}
L_d =id\sqrt{-g}\bar{\nu}_R D^2 \slashed{D}\nu_L= id \Bigg \lbrace & \bar{\tilde{\nu}}_R\left[ - h \partial^2  +  \partial_\mu h \partial^\mu  - \frac{i}{2} \sigma_{ab} \partial^b h \partial^a \right] \slashed{\partial}\tilde{\nu}_L \nonumber\\
& +  \bar{\tilde{\nu}}_R \partial^2 \left[-\frac{1}{2}h\slashed{\partial} + \frac{n-1}{4}\slashed{\partial}h \right] \tilde{\nu}_L \nonumber\\
&  + \frac{5}{4}h\bar{\tilde{\nu}}_R \partial^2 \slashed{\partial}\tilde{\nu}_L  - \frac{3}{4}\bar{\tilde{\nu}}_R\partial^2 \slashed{\partial} (h \tilde{\nu}_L) \Bigg \rbrace
\end{align}

\section{Discrete symmetries}

In this appendix, we evaluate the transformations under C, P and T
symmetries of the operators appearing in the effective action. Note
that we omit any total derivatives in the equations shown
below since they do not contribute to the action.

\subsection*{Charge Conjugation}
In what follows we make liberal use of the identity
\begin{equation}
\mathcal{C}\Gamma_\mu^T \mathcal{C} = \Gamma_\mu,
\end{equation}
as well as integration by parts and the property $C^{T}=-C$. We begin with the operator $L_a$ which transforms under C as follows:
\begin{align}
\mathcal{C}i \bar{\nu}_R \left[2R_{\mu \nu}\gamma^\mu D^\nu  +
  \frac{1}{2}\partial_\mu R \gamma^\mu \right] 
\nu_L \mathcal{C}^{-1}& = \left(- \nu^T C^{-1}\right) i \left[2R_{\mu \nu}\gamma^\mu D^\nu  + \frac{1}{2}\partial_\mu R \gamma^\mu \right] \frac{1-\gamma_5}{2}   C \left( \bar{\nu}\right)^T \nonumber\\
&= \bar{\nu}  i C \frac{1-\gamma_5^{T}}{2} \left[2R_{\mu \nu}\left(\overleftarrow{D}^\nu\right)^{T}   + \frac{1}{2}\partial_\mu R \right] \left(\gamma^\mu \right)^{T}   C^{-1}  \nu \nonumber\\
& = - \bar{\nu}  i \frac{1-\gamma_5}{2}   \left[2R_{\mu \nu} \overleftarrow{D}^\nu   + \frac{1}{2}\partial_\mu R \right] \gamma^\mu    \nu \nonumber\\
& = \bar{\nu}  i \frac{1-\gamma_5}{2}   \left[2R_{\mu \nu} D^\nu +  2
  \left( \nabla^\mu R_{\mu \nu}\right)   - \frac{1}{2}\partial_\mu R
\right] \gamma^\mu   \nu \nonumber\\ 
& = \bar{\nu}  i    \left[2R_{\mu \nu} D^\nu + \frac{1}{2}\partial_\mu R \right] \gamma^\mu \frac{1+\gamma_5}{2}   \nu \nonumber\\
& =\bar{\nu}_L  i    \left[2R_{\mu \nu} D^\nu +
  \frac{1}{2}\partial_\mu R \right] \gamma^\mu \nu_R \ .
\end{align}
The steps for the operator $L_c$ are identical and the operator $L_d$ follows in much the same way, with the additional use of the identity (\ref{dd2}). For the operator $L_b$ we have
\begin{align}
\mathcal{C} \left( \partial_\mu R \bar{\nu}_R \gamma^\mu \nu_L \right) \mathcal{C}^{-1} & = \partial_\mu R \left(- \nu^T C^{-1}\right) \gamma^\mu  \frac{1-\gamma_5}{2}  C \left( \bar{\nu}\right)^T \nonumber\\ 
 & =\partial_\mu R  \bar{\nu} C^T \frac{1-\gamma_5^T}{2} \left( \gamma^\mu\right)^T \left(C^{-1}\right)^T \nu \nonumber\\
 & =\partial_\mu R  \bar{\nu} C  \frac{1-\gamma_5^T}{2}\nu \left( \gamma^\mu\right)^T  C^{-1} \nonumber\\
& =\partial_\mu R  \bar{\nu} C  \frac{1-\gamma_5^T}{2}C^{-1}C \left( \gamma^\mu\right)^T  C^{-1} \nu  \nonumber\\
& =- \partial_\mu R  \bar{\nu}   \frac{1-\gamma_5}{2}\gamma^\mu \nu \nonumber\\
& =-\partial_\mu R  \bar{\nu}_L  \gamma^\mu  \nu_R \ . 
\end{align}
This establishes the C transformation properties of the operators in the effective action. 

\subsection*{Parity}
As an example we look at the parity transformation of $L_a$.  We have, using $\mathcal{P}\nu \mathcal{P}^{-1}=\gamma^0 \nu$:
\begin{align}
& \mathcal{P}i \bar{\nu}_R \left[2R_{\mu \nu}\gamma^\mu D^\nu  + \frac{1}{2}\partial_\mu R \gamma^\mu \right] 
\nu_L \mathcal{P}^{-1} \nonumber\\
&= \bar{\nu} i\left[2 (-)_\mu (-)_\nu R_{\mu \nu} (-)^\nu D^\nu+
  \frac{1}{2} (-)_\mu \partial_\mu R  \right] (-)^\mu \gamma^\mu 
\frac{1+ \gamma_5}{2}   \nu \nonumber\\
& = \bar{\nu} i\left[2  R_{\mu \nu} D^\nu+  \frac{1}{2} \partial_\mu R
\right]  \gamma^\mu \frac{1+ \gamma_5}{2}   \nu \nonumber\\
& =  \bar{\nu}_L i\left[2  R_{\mu \nu} \gamma^\mu D^\nu+  \frac{1}{2}
  \partial_\mu R \gamma^\mu  \right]   \nu_R \ .
\end{align}
where we used the result $\Gamma^\mu \gamma^0 = (-)^\mu \gamma^0
\Gamma^\mu$ to commute the $\gamma^0$ to the left. 
The operators $L_c$ and $L_d$ follow in a similar way. Next we look at $L_b$:
\begin{align}
\mathcal{P} \left( \partial_\mu R \bar{\nu}_R \gamma^\mu \nu_L \right)
\mathcal{P}^{-1} & = (-)_\mu \partial_\mu R \nu^\dagger \gamma^\mu  
\frac{1-\gamma_5}{2}  \gamma^0 \nu \nonumber\\ 
 & =\partial_\mu R  \bar{\nu}  \gamma^\mu \frac{1+\gamma_5}{2}   \nu \nonumber\\
& = \partial_\mu R  \bar{\nu}_L  \gamma^\mu \nu_R \ .
\end{align}
\subsection*{Time Reversal}
Using $B^\dagger \Gamma^{\mu*} B = (-)^\mu \Gamma^\mu$ we find
\begin{align}
& \mathcal{T}i \bar{\nu}_R \left[2R_{\mu \nu}\gamma^\mu D^\nu  + \frac{1}{2}\partial_\mu R \gamma^\mu \right] 
\nu_L \mathcal{T}^{-1} \\
&= \bar{\nu} \left[2 (-)_\mu (-)_\nu R_{\mu \nu} (-)^\nu D^\nu+  \frac{1}{2} (-)_\mu \partial_\mu R  \right] (-)^\mu \gamma^\mu \frac{1- \gamma_5}{2}   \nu\\
& = \bar{\nu} \left[2  R_{\mu \nu} D^\nu+  \frac{1}{2} \partial_\mu R  \right]  \gamma^\mu \frac{1- \gamma_5}{2}   \nu\\
& = \bar{\nu}_R \left[2  R_{\mu \nu} \gamma^\mu D^\nu+
  \frac{1}{2} \partial_\mu R \gamma^\mu \right] \nu_L  \ .
\end{align}
with the results for $L_c$ and $L_d$ following in a similar fashion. For $L_b$ we have
\begin{align}
\mathcal{T} \left( \partial_\mu R \bar{\nu}_R \gamma^\mu \nu_L \right)
\mathcal{T}^{-1} & 
= - (-)_\mu \partial_\mu R \nu^\dagger B^\dagger \gamma^{0*} \gamma^{\mu *}  \frac{1-\gamma_5^*}{2} B \nu \nonumber\\ 
 & =- \partial_\mu R  \bar{\nu} \gamma^\mu \frac{1-\gamma_5}{2}   \nu \nonumber\\
& = - \partial_\mu R  \bar{\nu}_R \gamma^\mu \nu_L.
\end{align}


\begin{thebibliography}{9}

\bibitem{Sakharov}
A.~Sakharov, Pisma Zh.~Eksp.~Teor.~Fiz. 5, 32 (1967), reprinted in 
E.~W.~Kolb and M.~S.~Turner (eds.), \emph{The Early Universe},
Addison-Wesley, Reading, Massachusetts, 1988.

\bibitem{Cohen:1987vi}
  A.~G.~Cohen and D.~B.~Kaplan,
  Phys.\ Lett.\ B {\bf 199} (1987) 251.

\bibitem{Lambiase:2013haa}
  G.~Lambiase, S.~Mohanty and A.~R.~Prasanna,
  Int.\ J.\ Mod.\ Phys.\ D {\bf 22} (2013) 1330030
  [arXiv:1310.8459 [hep-ph]].

\bibitem{Davoudiasl:2004gf}
  H.~Davoudiasl, R.~Kitano, G.~D.~Kribs, H.~Murayama and P.~J.~Steinhardt,
  Phys.\ Rev.\ Lett.\  {\bf 93} (2004) 201301
  [hep-ph/0403019].

\bibitem{Drummond:1979pp}
  I.~T.~Drummond and S.~J.~Hathrell,
  Phys.\ Rev.\ D {\bf 22} (1980) 343.

\bibitem{Ohkuwa:1980jx}
  Y.~Ohkuwa,
  Prog.\ Theor.\ Phys.\  {\bf 65} (1981) 1058.

\bibitem{Colladay:1998fq}
  D.~Colladay and V.~A.~Kostelecky,
  Phys.\ Rev.\ D {\bf 58} (1998) 116002
  [hep-ph/9809521].

\bibitem{Kuzmin:1985mm}
  V.~A.~Kuzmin, V.~A.~Rubakov and M.~E.~Shaposhnikov,
  Phys.\ Lett.\ B {\bf 155} (1985) 36.

\bibitem{Singh:2003sp}
  P.~Singh and B.~Mukhopadhyay,
  Mod.\ Phys.\ Lett.\ A {\bf 18} (2003) 779.

\bibitem{Mukhopadhyay:2005gb}
  B.~Mukhopadhyay,
  Mod.\ Phys.\ Lett.\ A {\bf 20} (2005) 2145
  [astro-ph/0505460].

\bibitem{Debnath:2005wk}
  U.~Debnath, B.~Mukhopadhyay and N.~Dadhich,
  Mod.\ Phys.\ Lett.\ A {\bf 21} (2006) 399
  [hep-ph/0510351].

\bibitem{Bir}
N.~D.~Birrel and P.~C.~W.~Davies, \emph{Quantum Fields in Curved
  Space}, Cambridge University Press, 1982.

\bibitem{Free}
D.~Z.~Freedman and A.~Van Proeyen, \emph{Supergravity},
Cambridge University Press, 2012.

\bibitem{Poisson:2011nh}
  E.~Poisson, A.~Pound and I.~Vega,
  Living Rev.\ Rel.\  {\bf 14} (2011) 7
  [arXiv:1102.0529 [gr-qc]].

\bibitem{MTW}
C.~W.~Misner, K.~S.~Thorne and J.~A.~Wheeler, \emph{Gravitation},
Freeman, San Francisco, 1973.

\bibitem{Schneider}
P.~Schneider, J.~Ehlers and E.~E.~Falco, \emph{Gravitational Lenses},
Springer-Verlag, Berlin, 1992.

\bibitem{Shore:2003jx}
  G.~M.~Shore,
  ``\emph{Causality and superluminal light},''
  in ``Time and Matter: Proceedings of the International
  Colloquium on the Science of Time'', Venice 2002, I.~I.~Bigi and
  M.~Faessler (eds.), World Scientific, Singapore, 2006. ~~
  gr-qc/0302116.

\bibitem{Audretsch:1981wf}
  J.~Audretsch,
  J.\ Phys.\ A {\bf 14} (1981) 411.

\bibitem{Lambiase:2006md}
  G.~Lambiase and S.~Mohanty,
  JCAP {\bf 0712} (2007) 008
  [astro-ph/0611905].

\bibitem{Lambiase:2011by}
  G.~Lambiase and S.~Mohanty,
  Phys.\ Rev.\ D {\bf 84} (2011) 023509
  [arXiv:1107.1213 [hep-ph]].

\bibitem{Ellis:2013gca}
  J.~Ellis, N.~E.~Mavromatos and S.~Sarkar,
  Phys.\ Lett.\ B {\bf 725} (2013) 407
  [arXiv:1304.5433 [gr-qc]].

\bibitem{Hollowood:2010bd}
  T.~J.~Hollowood and G.~M.~Shore,
  Phys.\ Lett.\ B {\bf 691} (2010) 279
  [arXiv:1006.0145 [hep-th]].

\bibitem{Hollowood:2011yh}
  T.~J.~Hollowood and G.~M.~Shore,
  JHEP {\bf 1202} (2012) 120
  [arXiv:1111.3174 [hep-th]].

\bibitem{Shore:2007um}
  G.~M.~Shore,
  Nucl.\ Phys.\ B {\bf 778} (2007) 219
  [hep-th/0701185].

\bibitem{Hollowood:2007ku}
  T.~J.~Hollowood and G.~M.~Shore,
  Nucl.\ Phys.\ B {\bf 795} (2008) 138
  [arXiv:0707.2303 [hep-th]].

\bibitem{Hollowood:2008kq}
  T.~J.~Hollowood and G.~M.~Shore,
  JHEP {\bf 0812} (2008) 091
  [arXiv:0806.1019 [hep-th]].



\end{thebibliography}
\end{document}